\documentclass[usenatbib]{mn2e}
\usepackage{hyperref}
\hypersetup{colorlinks,citecolor=Blue,linkcolor=Red,urlcolor=Blue}
\usepackage{amssymb}
\usepackage{amsmath}
\usepackage{empheq}
\usepackage{mathrsfs}
\usepackage[usenames,dvipsnames]{color}
\allowdisplaybreaks 

\newcommand{\G}{\mathcal{G}}
\newcommand{\M}{M_{\star}}
\newcommand{\Ham}{\tilde{\mathcal{H}}}
\newcommand{\PSI}{\tilde{\Psi}}
\newcommand{\past}{p^{\ast}_{\rm{eq}}}

\newcommand{\Poincare}{{Poincar$\acute{\rm{e}}$}}
\newcommand{\appropto}{\mathrel{\vcenter{\offinterlineskip\halign{\hfil$##$\cr\propto\cr\noalign{\kern2pt}\sim\cr\noalign{\kern-2pt}}}}}

\begin{document}
 
\title[MMR Capture Theory]{Capture of Planets Into Mean Motion Resonances and the Origins of Extrasolar Orbital Architectures}

\author[K. Batygin]{Konstantin Batygin\thanks{kbatygin@gps.caltech.edu}
\newauthor \\
Division of Geological and Planetary Sciences, California Institute of Technology, 1200 E. California Blvd., Pasadena, CA 91125, USA}

\maketitle
 
\begin{abstract}
The early stages of dynamical evolution of planetary systems are often shaped by dissipative processes that drive orbital migration. In multi-planet systems, convergent amassing of orbits inevitably leads to encounters with rational period ratios, which may result in establishment of mean motion resonances. The success or failure of resonant capture yields exceedingly different subsequent evolutions, and thus plays a central role in determining the ensuing orbital architecture of planetary systems. In this work, we employ an integrable Hamiltonian formalism for first order planetary resonances that allows both secondary bodies to have finite masses and eccentricities, and construct a comprehensive theory for resonant capture. Particularly, we derive conditions under which orbital evolution lies within the adiabatic regime, and provide a generalized criterion for guaranteed resonant locking as well as a procedure for calculating capture probabilities when capture is not certain. Subsequently, we utilize the developed analytical model to examine the evolution of Jupiter and Saturn within the protosolar nebula, and investigate the origins of the dominantly non-resonant orbital distribution of sub-Jovian extrasolar planets. Our calculations show that the commonly observed extrasolar orbital structure can be understood if planet pairs encounter mean motion commensurabilities on slightly eccentric ($e\sim0.02$) orbits. Accordingly, we speculate that resonant capture among low-mass planets is typically rendered unsuccessful due to subtle axial asymmetries inherent to the global structure of protoplanetary disks.
\end{abstract}

\begin{keywords}
planets and satellites: dynamical evolution and stability, celestial mechanics, methods: analytical
\end{keywords}

\section{Introduction}
The orbital architectures of the Solar System's planetary and satellite populations, as well as the currently known aggregate of extrasolar planets, exhibit numerous peculiar features. In principle, each specific attribute of the galactic planetary census entails delicate constraints that inform the dominant mechanisms responsible for its inception and subsequent evolution. Therefore, theoretical interpretation of the observed state of planetary systems ultimately holds the key to understanding their origins.

Among the most striking characteristics inherent to the orbital distribution of planets and satellites is the modest predisposition for orbital resonances, or loosely speaking, pairs of orbits with nearly rational periods. In the Solar System, the preference for commensurability among satellite pairs (over randomly distributed orbits) is well-recognized, and was first pointed out more than half a century ago by \cite{RoyOvenden1954} (see also \citealt{Goldreich1965} and \citealt{Dermott1973}). 

In the extrasolar realm, the radial-velocity sub-sample points to a mild over-abundance of resonant giant planets, typically at orbital radii in excess of $\sim1$ AU \citep{Wright2011,WinnFabrycky2014}. Meanwhile, the period-ratio distribution of an extensive collection of sub-Jovian planets discovered by \textit{Kepler} and other surveys exhibits notable enhancements at values slightly outside of exact 2:1 and 3:2 commensurabilities \citep{Fabrycky2014}. Instances of multi-resonant chains, akin to the Galilean satellites, have also been found\footnote{Examples of such systems include GJ\,876, Kepler-79 and Kepler-223.}. 
%Despite these features, however, the dominant majority of the observational sample is not resonant.

Collectively, these observations imply that planet formation environments can be congenial towards emergence of resonant systems. However, the overwhelming majority of observed commensurabilities are unlikely to be truly primordial in nature. Instead, resonances generally result from dissipative convergent evolution of pairs of orbits that follows the initial phase of conglomeration \citep{Goldreich1965,Allan1969,Allan1970,Sinclair1970,Sinclair1972}. 

There exists a multitude of physical mechanisms that may cause orbits to slowly approach one-another. Specifically, migration facilitated by tidal dissipation \citep{GoldreichSoter1966, Peale1976, Peale1986, YoderPeale1981}, interactions with a gaseous disk \citep{GoldreichTremaine1980,Crida2008,KleyNelson2012}, and scattering of debris within a planetesimal swarm \citep{FernandezIp1984,Malhotra1993,Malhotra1995,Kirsh2009} are the most frequently quoted transport processes relevant to planets and satellites.

While convergent migration is required for resonances to congregate, this process alone does not guarantee successful resonant locking. In fact, the outcome of an encounter with a mean-motion commensurability depends on the intrinsic parameters of the system, such as the planet-star and planet-planet mass ratios, the convergence rate, as well as the degree of pre-encounter orbital excitation. However, in spite of the uncertainty in the prospect of resonant capture itself, the post-encounter evolution of the system may depend critically on its result. 

As an example of resonances' decisive nature, consider the fate of a pair of giant planets, at an epoch when the gaseous protoplanetary disk has not yet dispersed. Although individually the planetary orbits would decay towards the central star along with the accretionary flow of the gas, successful capture of such objects into mean-motion resonance can reverse the overall inward drift and allow the planets to remain at large orbital radii\footnote{Indeed, it is believed that this exact process is responsible for the retention of Jupiter and Saturn in the outer regions of the Solar System \citep{MorbyCrida2007,Pierens2014}.} \citep{MassetSnellgrove2001,DAngeloMarzari2012}. As another example, assembly of sub-Jovian planets into resonances facilitates stabilization by the phase-protection mechanism \citep{Greenberg1977} and allows chains of planets to coherently migrate inwards without encountering each-other\footnote{The long-term post-nebular stability of such systems is a separate, non-trivial issue \citep{Chambers1996,MahajanWu2014}.} (\citealt{TerquemPapaloizou2007, CresswellNelson2008}; see also \citealt{LeePeale2002}). On the contrary, failure of objects to capture into resonance leads to an impulsive change in the system's orbital properties \citep{TittemoreWisdom1988,TittemoreWisdom1989,TittemoreWisdom1990} and may trigger large-scale dynamical instabilities \citep{Tsiganis2005,Morbyetal2007,BatyginBrown2010}. \textit{Thus, quantification of the capture probability of planets into mean motion resonance is central to understanding the early evolution of planetary systems.}

Over the last half century, numerous efforts have been made to develop a complete understanding of resonant capture. Following the initial studies of \citet{Goldreich1965,GoldreichSoter1966,Allan1969,Allan1970,Sinclair1972,Greenberg1973}, the first analytical formulation of the capture probability within the framework of the \textit{circular restricted three-body problem} (i.e. where a massive planet is assumed to orbit the central star on a circular trajectory and perturb a massless test-particle) was proposed\footnote{It should be noted that the concept of probabilistic capture within the framework of spin-orbit resonances was first introduced by \citet{GoldreichPeale1966}.} by \citet{Yoder1973,Yoder1979} and independently by \citet{Neishtadt1975}. Fully employing the use of adiabatic invariants, the formalism for analytic determination of capture probabilities was generalized by \citet{Henrard1982} (see also \citealt{HenrardLemaitre1983,Lemaitre1984}) and subsequently simplified by \citet{BorderiesGoldreich1984} (see also \citealt{Peale1986,Malhotra1988}). Additional studies aimed at extending resonant capture theory to the non-adiabatic regime have since also been undertaken by \citet{Friedland2001,Quillen2006,Ketchum2011} and \citet{OgiharaKobayashi2013}.

While the paradigm of the circular restricted three-body problem lends itself easily to theoretical analysis, it is not directly applicable to the ultimately relevant issue of resonant capture of massive planets with non-circular orbits. Accordingly, to attack this more complicated problem, numerous authors have resorted to the use of numerical experiments. To this end, simulations that mimic the dissipative effects with fictitious forces \citep{LeePeale2002,HahnMalhotra2005,TerquemPapaloizou2007,ReinPapaloizou2010} as well as self-consistent hydrodynamical \citep{Kley2005,PapaloizouSzuszkiewicz2005,Crida2008,CresswellNelson2008}, self-gravitational \citep{Moore2008,MooreQuillen2011} and tidal calculations \citep{Mardling2008,LithwickWu2012,BatyginMorby2013} have been performed. 

Although the published simulations represent important advancements in the understanding of planet formation and evolution, such calculations continue to leave analytical development to be desired for two reasons. First, numerical experiments are typically tailored towards particular systems with specific choices of physical parameters. Without the knowledge of underlying theoretical relationships, it becomes difficult to translate the obtained results to another case without additional modeling. Second, despite substantial advances in computational technologies, high-resolution numerical experiments remain much too computationally expensive to practically serve as a replacement for theory.

Consequently, in order to explain why planetary systems  possess the orbital architecture that they have, it is necessary to generalize the existing resonance capture theory to the physical domain of the \textit{unrestricted elliptic three-body problem}. This is the primary purpose of this work. Ultimately, with a more complete model in hand, we shall delineate parameter regimes within which resonant capture is expected to be successful, and aim to understand why resonant orbits are neither ubiquitous nor absent within the galactic planetary census. 

The paper is structured as follows. In section 2, we generalize the existing first order resonance capture theory to allow for both planets to have masses as well as finite eccentricities. Additionally, we obtain a criterion which dictates whether or not the dissipative evolution lies in the adiabatic regime. In section 3, we examine some immediate consequences of the model. Particularly, we construct maps of capture probabilities relevant to adiabatic encounters and derive specific adiabatic thresholds for disk-driven migration, planetesimal driven migration, and tidal evolution. Subsequently, we analytically consider the orbital evolution of Jupiter and Saturn while submerged within the protosolar nebula, and derive critical eccentricities above which sub-Jovian extrasolar planets fail to capture into mean motion resonances. We summarize and discuss our results in section 4.

\section{Analytical Theory}

The calculation we aim to perform is perturbative in nature. Accordingly, we begin with a basic description of the dynamics. It is standard practice in celestial mechanics to treat planet-planet interactions as perturbations to Keplerian motion. As such, planet-planet potential is typically expanded as a Fourier series in the orbital angles and a power series in the planetary eccentricities and inclinations \citep{LaskarRobutel1995,EllisMurray2000,LaskarBoue2010}. 

In the vicinity of a given resonance, it is sensible to average over all short-periodic terms (i.e. those that vary on an orbital timescale) in the expansion, and retain only slowly varying (compared to the orbital period) harmonics, which include those associated with the resonant interaction \citep{MD99}. This procedure yields a simplified Hamiltonian that approximates the full Hamiltonian near a commensurability.

\subsection{Model Hamiltonian}

\begin{figure}
\includegraphics[width=1\columnwidth]{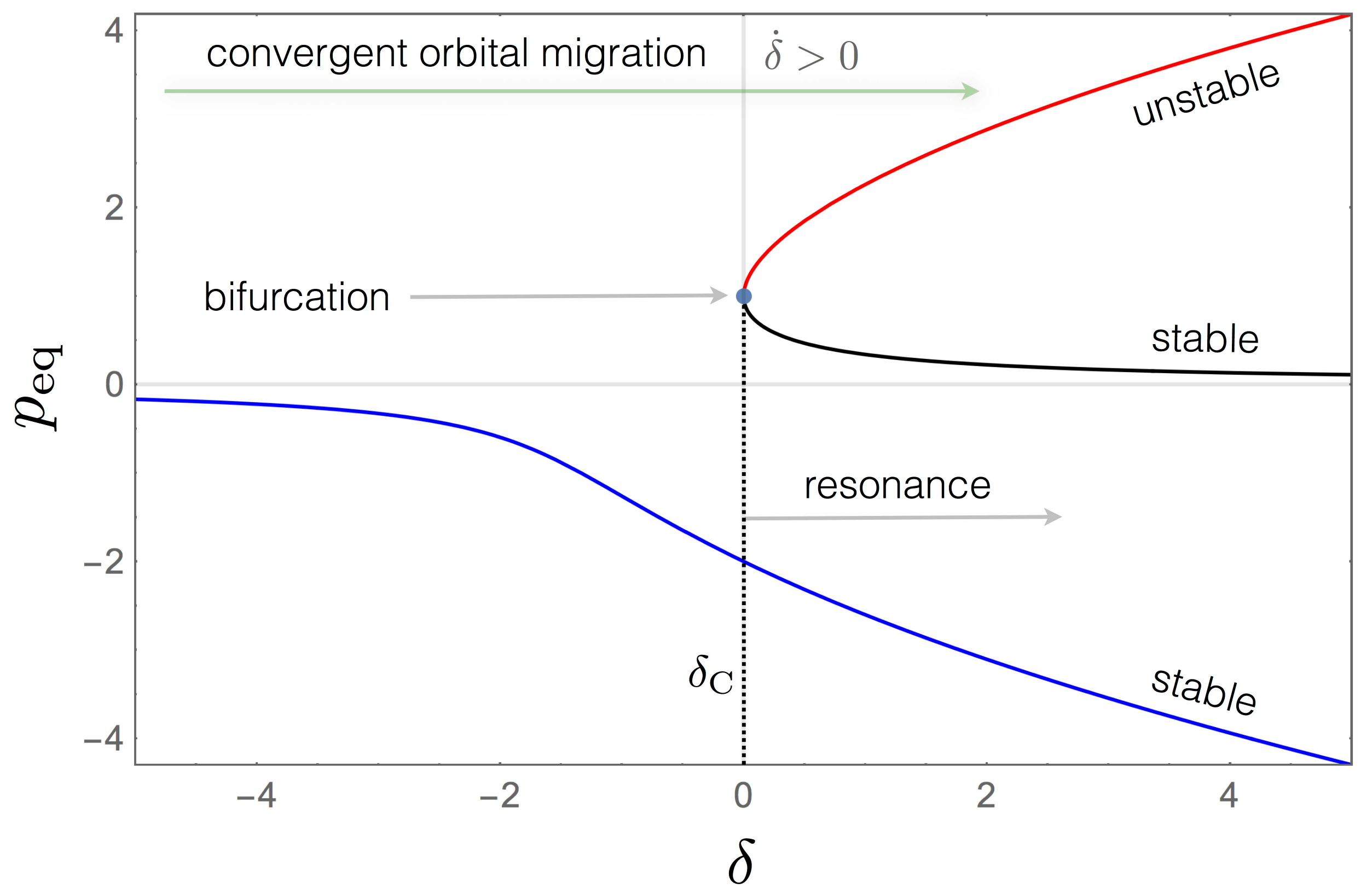}
\caption{Equilibria of Hamiltonian (\ref{SFMR}). The three roots of equation (\ref{equilibrium}) are shown as functions of the resonance proximity parameter. The elliptic equilibrium point that is always real and negative is shown in blue. The elliptic and hyperbolic fixed points that only exist above the bifurcation value of $\delta \geqslant \delta_{\rm{C}}$ are shown in black and red respectively. Extrinsically driven convergent evolution of the orbits corresponds to a gradual increase in $\delta$.}
\label{equlib}
\end{figure}

In its most rudimentary form (i.e. to first order in planetary masses and eccentricities), the planar $k:k-1$ resonant Hamiltonian reads:
\begin{align}
\label{Horbel}
\mathcal{H} = &- \frac{\mathcal{G} \M m_1}{2 a_1} - \frac{\mathcal{G} \M m_2}{2 a_2} - \frac{\mathcal{G} m_1 m_2}{a_2} \nonumber \\
&\times \bigg{[} f_{\rm{res}}^{(1)} e_1 \cos(k \lambda_2 - (k-1) \lambda_1 - \varpi_1)  \nonumber \\
&+ f_{\rm{res}}^{(2)} e_2 \cos(k \lambda_2 - (k-1) \lambda_1 - \varpi_2) \bigg{]}.
\end{align}
In the above expression, $k$ is an integer greater than unity, $\mathcal{G}$ is the gravitational constant,  $\M$ is the mass of the central star, $m$ is the planetary mass, $a$ is the semi-major axis, $e$ is the eccentricity, $\lambda$ is the mean longitude, $\varpi$ is the longitude of perihelion, and $f_{\rm{res}}$'s are functions of order unity that (weakly) depend on the semi-major axis ratio $(a_1/a_2)$ (see \citealt{CallegariYokoyama2007} for explicit expressions). The subscripts $1$ and $2$ denote the inner and outer planets respectively.

Upon inspection it is immediately clear that the Hamiltonian (\ref{Horbel}) contains two resonant arguments that appear in two separate harmonics. Although at first glance this Hamiltonian appears to comprise a typical two degree of freedom system, it was shown to be integrable by \citet{Sessin1981,SessinFerraz-Mello1984}. The published calculations of \citet{SessinFerraz-Mello1984} are rather involved, as they were performed within the framework of the \citet{Hori1966} perturbation method. However, the demonstration of integrability was greatly simplified by \citet{Wisdom1986} and \citet{Henrard1986}, who showed that instead of constructing formal perturbation series, it is possible to consolidate the two harmonics into a single term with the aid of a reducing canonical transformation\footnote{Evidently, a transformation of this sort was first proposed by \citet{Poincare1899}.} that corresponds to a rotation in phase-space (see also \citealt{HenrardLemaitre2005}). 

By employing the reducing transformation, we shall cast the Hamiltonian into a form that is synonymous to a first-order Andoyer Hamiltonian, also known as the second fundamental model for resonance \citep{HenrardLemaitre1983,Morbidelli2002}. In this work, we will not spell out the derivation in excruciating detail, as it is available else-where \citep{SessinFerraz-Mello1984,Wisdom1986,Henrard1986,Ferraz-Mello2007,BatyginMorbidelli2013AA,Deck2013,Delisle2014}. Instead, we shall restrict ourselves to sketching out the important steps. In both notation and substance, we shall closely follow the derivation outlined by \citet{BatyginMorbidelli2013AA}. 

\begin{figure*}
\includegraphics[width=1\textwidth]{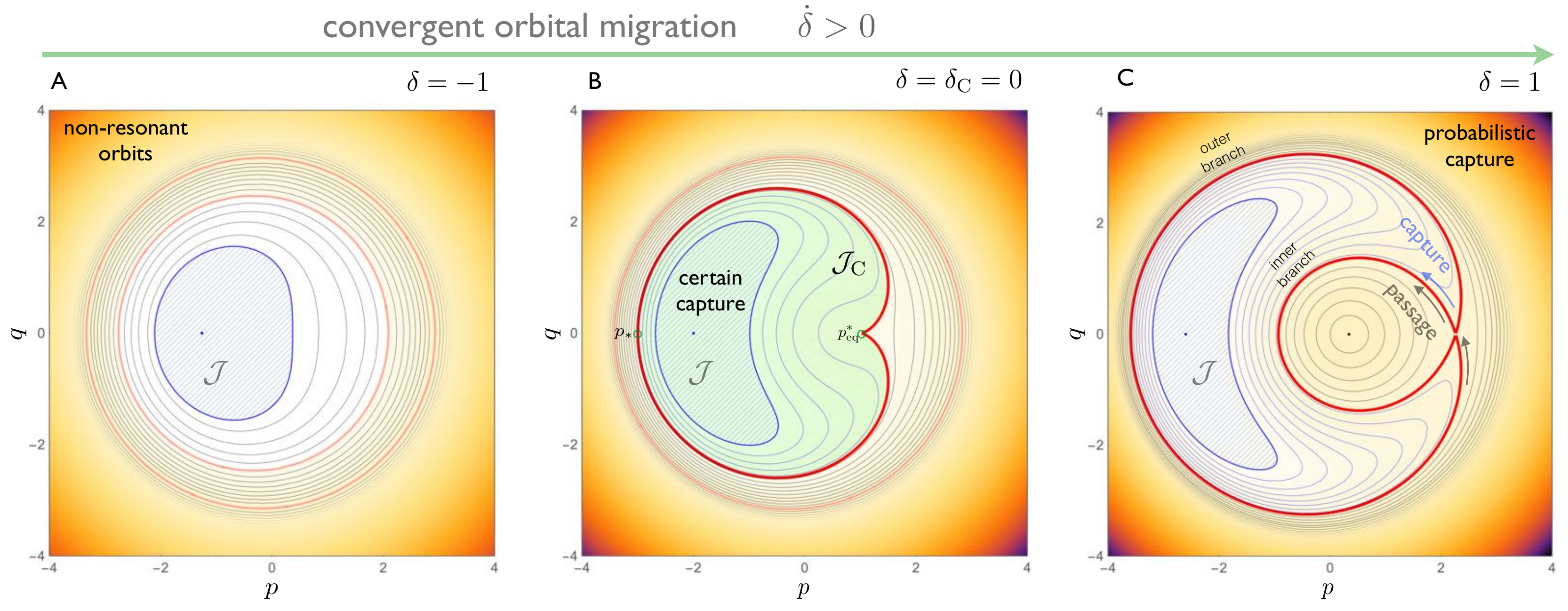}
\caption{Evolution of the phase-space portrait of the Hamiltonian during convergent orbital migration. Panels A, B and C depict level curves of equation (\ref{SFMR}) for $\delta = -1$, $0$, and $1$ respectively. The equilibrium points are shown with small dots and are color-coded in the same way as in Figure (\ref{equlib}). While libration of the critical angle is possible in panel A, formal resonant trajectories only exist in panels B and C, where the separatrix is shown as a thick red curve. Across the panels, three values of the phase space area (adiabatic invariant) are accentuated: sub-critical (labeled $\mathcal{J}$ and shown as a shaded blue region), critical (labeled $\mathcal{J}_{\rm{C}}$ and emphasized with a green shade in panel B), and super-critical (corresponding to the area occupied by the outer branch of the separatrix on panel C). Trajectories encircling the critical and super-critical areas are shown on panel A as thickened red lines. A circulating orbit engulfing a super-critical area is similarly shown in panel B. While smooth adiabatic entry into resonance is possible for initially circulating trajectories occupying a phase-space area equal to, or smaller than $\mathcal{J}_{\rm{C}}$, those occupying a larger area must cross the separatrix to be captured, as shown on panel C. On panel B, the minimal and maximal excursions of the action along the separatrix (which correspond to the resonance width at the onset of resonance) are labled $\past$ and $p_{\ast}$ respectively.}
\label{portraits}
\end{figure*}

As a first step, we define the canonical variables
\begin{align}
\label{transformone}
\mathcal{K} &= \Lambda_1 + \frac{k -1}{k} \Lambda_2  &  \kappa = \lambda_1& \nonumber \\
\Theta &= \Lambda_2 / k  &  \theta = k \lambda_2 - (k-1) \lambda_1& \nonumber \\
x_i &= e_i \sqrt{ [\Lambda]_i } \cos(-\varpi_i) &  y_i = e_i \sqrt{ [\Lambda]_i } \sin(-\varpi_i)&
\end{align}
where $\Lambda_i = m_i \sqrt{\G \M a_i}$ and the square brackets $[\ ]$ denote an evaluation at nominal resonant semi-major axis (i.e. $[a]_2 = (k/(k-1))^{2/3} [a]_1$). Upon expanding the Keplerian part of the Hamiltonian to second order in $\Lambda$ around $[\Lambda]$ (which corresponds to a perturbation that is first order in $e$), neglecting semi-major axis variation in the disturbing part of the Hamiltonian (since it is already first order in $e$), and dropping constant terms, the expression (\ref{Horbel}) takes the form:
\begin{align}
\label{Htheta}
\mathcal{H} &= 3[h]_1 (k-1) \mathcal{K} \Theta  -\frac{3}{2}\big([h]_1(k-1)^2 + [h]_2 k^2  \big) \Theta^2 \nonumber \\
&- (\alpha x_1 + \beta x_2) \cos(\theta) + (\alpha y_1 + \beta y_2) \sin(\theta).
\end{align}
Here, $[h]= [n]/[\Lambda] = 1/(m [a]^2)$ is the inverse moment of inertia of a circular orbit, and
\begin{align}
\label{alpha}
\alpha &=  \frac{\mathcal{G}^2 \M m_1 m_2^3}{[\Lambda]_2^2} \frac{ f_{\rm{res}}^{(1)}}{\sqrt{[\Lambda]_{1}}} \nonumber \\
\beta &= \frac{\mathcal{G}^2 \M m_1 m_2^3}{[\Lambda]_2^2} \frac{ f_{\rm{res}}^{(2)}}{\sqrt{[\Lambda]_{2}}}
\end{align}
are constants that encapsulate the strengths of the individual harmonics present in the Hamiltonian. 

\begin{figure*}
\includegraphics[width=1\textwidth]{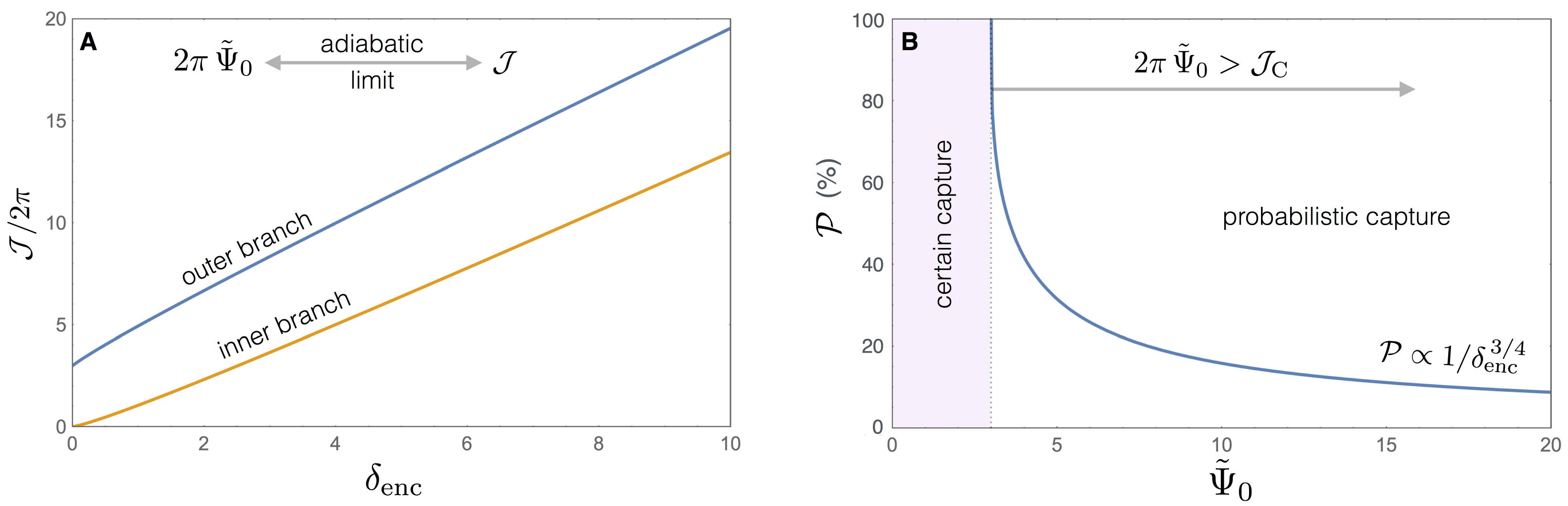}
\caption{Area index and capture probability. Panel A shows the phase-space area occupied by the inner and outer branches of the separatrix as a function of the proximity parameter. In the adiabatic regime, the initial phase-space area occupied by a circulating trajectory far away from resonance matches onto the area of the outer branch of the separatrix, evaluated at the proximity parameter, $\delta_{\rm{enc}}$, at which the resonant encounter occurs. Panel B shows the resonance capture probability as a function of the initial action. Note that the capture probability diminishes rapidly for values of the initial action that exceed the critical value below which capture is certain.}
\label{pcapt}
\end{figure*}

Hamiltonian (\ref{Htheta}) is independent of the angle $\kappa$, meaning that the action $\mathcal{K}$ is a constant of motion. However, the number of degrees of freedom is still too great for integrability. Defining a canonical rotation \citep{Wisdom1986,Henrard1986,Ferraz-Mello2007}
\begin{align}
\label{suicide}
u_1 &= \frac{\alpha x_1 + \beta x_2}{\sqrt{\alpha^2+\beta^2}} &   v_1 = \frac{\alpha y_1 + \beta y_2}{\sqrt{\alpha^2+\beta^2}}& \nonumber \\
u_2 &= \frac{\beta x_1 - \alpha x_2}{\sqrt{\alpha^2+\beta^2}} &  v_2 = \frac{\beta y_1 - \alpha y_2}{\sqrt{\alpha^2+\beta^2}}&
\end{align}
and the associated action-angle coordinates
\begin{align}
\label{suicidepolar}
u_i &= \sqrt{2 \Phi_i} \cos (\phi_i) & v = \sqrt{2 \Phi_i} \sin (\phi_i)&,
\end{align}
the Hamiltonian is reduced to a form that contains only a single harmonic: 
\begin{align}
\label{Htheta2}
\mathcal{H} &= 3[h]_1 (k-1) \mathcal{K} \Theta  -\frac{3}{2}\big([h]_1(k-1)^2 + [h]_2 k^2  \big) \Theta^2 \nonumber \\
&  - \sqrt{\alpha^2+\beta^2} \sqrt{2 \Phi_1} \cos(\phi_1 + \theta).
\end{align}
This reduction identifies a second integral of motion, that is $\Phi_2$.

Employing a contact transformation, we define a final set of action-angle variables
\begin{align}
\label{finalactionangle}
\Psi &= \Phi_1 & \psi = \phi_1 + \theta& \nonumber \\
\Omega &= \Theta - \Psi & \sigma = \theta&
\end{align}
which renders $\mathcal{H}$ independent of $\sigma$. Accordingly, identifying $\Omega$ as the final conserved quantity and dropping constant terms, the Hamiltonian is cast into an integrable form:
\begin{align}
\label{Hinteg}
\mathcal{H} &= -3\left([h]_2 k^2 \Omega -[h]_1 (k-1) \left( \mathcal{K} - (k-1) \Omega \right) \right) \Psi   \nonumber \\
&-\frac{3}{2}\big([h]_1(k-1)^2 + [h]_2 k^2  \big) \Psi^2  \nonumber \\
&- \sqrt{\alpha^2+\beta^2} \sqrt{2 \Psi} \cos(\psi).
\end{align}

The constants that precede each of the terms in the Hamiltonian are not truly independent. Thus, as a concluding step in the preparation of the Hamiltonian, we shall scale the action and time, in order to introduce a single \textit{resonance proximity parameter}, $\delta$. To maintain symplecticity, we scale all of the actions present in the Hamiltonian, as well as the Hamiltonian itself by the same constant factor, $\eta$:
\begin{align}
\label{scale}
\tilde{\mathcal{H}} = \mathcal{H}/\eta && \tilde{\Psi} = \Psi/\eta &&  \tilde{\mathcal{K}} = \mathcal{K}/\eta && \tilde{\Omega} = \Omega/\eta.
\end{align}
We choose the expression for $\eta$ such that the coefficient upfront the term that is quadratic in action is half\footnote{This is a slightly different scaling from that employed by \citet{BatyginMorbidelli2013AA}.} as big as the one that precedes the harmonic:
\begin{align}
\label{eta}
\eta &= \left( \frac{(\alpha^2 + \beta^2)}{9([h]_1 (k-1)^2 + [h]_2 k^2)^2} \right)^{1/3} = m_1 m_2 \sqrt{\mathcal{G} \M [a]_2} \nonumber \\ 
&\times \left [\frac{ (f_{\rm{res}}^{(2)})^2 m_1 + (f_{\rm{res}}^{(1)})^2 (k/(k-1))^{1/3} m_2}{9 \M^2 \left(k^2 m_1 + (k-1)^{2/3} k^{4/3} m_2 \right)^2} \right]^{1/3}.
\end{align}
Subsequently, we change the unit of time to $3 \eta ([h]_1 (k-1)^2 + [h]_2 k^2) /2$, and divide $\mathcal{H}$ by the same factor. Writing the proximity parameter as 
\begin{align}
\label{delta}
\delta &= -\big( [h]_2 k^2 (3+2 \tilde{\Omega}) + [h]_1 (k-1) \nonumber \\
 &\times ( 3 k - 2 \tilde{\mathcal{K}} +2 (k-1) \tilde{\Omega} -3 ) \big)  \nonumber \\
 &\times \big([h]_1 (k-1)^2 + [h]_2 k^2\big)^{-1},
\end{align}
the Hamiltonian takes on the familiar form:
\begin{empheq}[box=\fbox]{align}
\label{SFMR}
\Ham = 3 (\delta+1) \PSI - \PSI^2 - 2 \sqrt{2 \PSI} \cos(\psi).
\end{empheq}

With the exception of the signs\footnote{Signs only determine if the resonant equilibrium point resides at $\psi = 0$ or $\psi = \pi$, and are thus unimportant to the discussion at hand.}, this is the Hamiltonian considered by \citet{Yoder1973,Yoder1979,Neishtadt1975,Henrard1982} and \citet{BorderiesGoldreich1984}. Note that this expression for the Hamiltonian is written in terms of dimensionless variables. Thus, the quantity $\eta$ (which has units of angular momentum) encapsulates all of the information regarding how the dynamics scales with mass ratios and physical sizes of the orbits. 

%Consequently, from here we can simply proceed by following previously established results. 
\subsection{Equilibria of the Hamiltonian}

Some insight into the properties of the Hamiltonian can be obtained by considering its equilibria. Defining the cartesian coordinates 
\begin{align}
\label{cart}
p =  \sqrt{2 \PSI} \cos(\psi)  \ \ \ \ \ q =  \sqrt{2 \PSI} \sin(\psi),
\end{align}
the Hamiltonian can be written as:
\begin{align}
\label{H}
\Ham = 3(\delta+1) \left(\frac{p^2 +q^2}{2} \right) - \left(\frac{p^2 +q^2}{2} \right)^2- 2 p.
\end{align}
It can be understood from equation (\ref{H}) that the equilibrium points must reside on the $p-$axis, since $\psi=0, \pi$ for $\partial \Ham/ \partial \psi = 0$. Thus, setting $q = 0$, the equilibrium equation reads:
\begin{align}
\label{equilibrium}
3(\delta+1) p - p^3 - 2 = 0.
\end{align}

The cubic equation (\ref{equilibrium}) admits three roots, one of which is always real and negative, while the other two are real (and positive) only above a critical bifurcation value of $\delta$, which can be computed by setting the two positive roots equal to each other. For Hamiltonian (\ref{SFMR}), the bifurcation occurs at $\delta_{\rm{C}} = 0.$ Strictly speaking, $\delta \geqslant \delta_{\rm{C}}$ defines a condition for the existence of resonance, since a homoclinic curve (i.e. separatrix), necessary for identification of resonant trajectories, only exists for values of $\delta$ greater than or equal to $\delta_{\rm{C}}$ (see e.g. \citealt{Delisle2012}). To this end, it is worth noting that although the functional form for $\delta$ is complicated, it monotonically grows as orbits approach each-other convergently. Thus, an increase in the value of $\delta$ unilaterally signals evolution into the resonance.

The three roots of equation (\ref{equilibrium}) are shown as functions of $\delta$ in Figure (\ref{equlib}) and can be classified as follows. The negative root (shown with a blue line) always corresponds to a stable (elliptical) equilibrium point and for $\delta \geqslant \delta_{\rm{C}}$, is enveloped by resonant trajectories. The positive root, whose value is closer to zero (shown with a black line), also corresponds to a stable fixed point but lies at the center of the inner circulation region of the phase space portrait. The larger positive root (shown with a red line) corresponds to an unstable (hyperbolic) equilibrium. For reference, phase-space portraits of the Hamiltonian for $\delta = -1$,  $\delta = \delta_{\rm{C}} = 0$ and $\delta = 1$ are depicted in Figure (\ref{portraits}), where the fixed points are color-coded in the same way as in Figure (\ref{equlib}).

Note that the unstable equilibrium defines a point where the inner and outer branches of the separatix join. For this reason, its value is particularly important for the evaluation of capture probabilities. Written explicitly, the corresponding expression for the unstable fixed point reads:
\begin{align}
\label{past}
\past&= \bigg(\imath \sqrt{3} \delta +\delta -\imath \sqrt{3} \left(\sqrt{-\delta  (\delta  (\delta +3)+3)}+1\right)^{2/3} \nonumber \\
&\times \left(\sqrt{-\delta  (\delta  (\delta +3)+3)}+1\right)^{2/3}+\imath \sqrt{3}+1\bigg) \nonumber \\
&\times \bigg(2 \sqrt[3]{\sqrt{-\delta  (\delta  (\delta +3)+3)}+1}\bigg)^{-1},
\end{align}
where $\imath = \sqrt{-1}$. Recall that this expression is purely real only for $\delta \geqslant 0$.

\begin{figure}
\includegraphics[width=1\columnwidth]{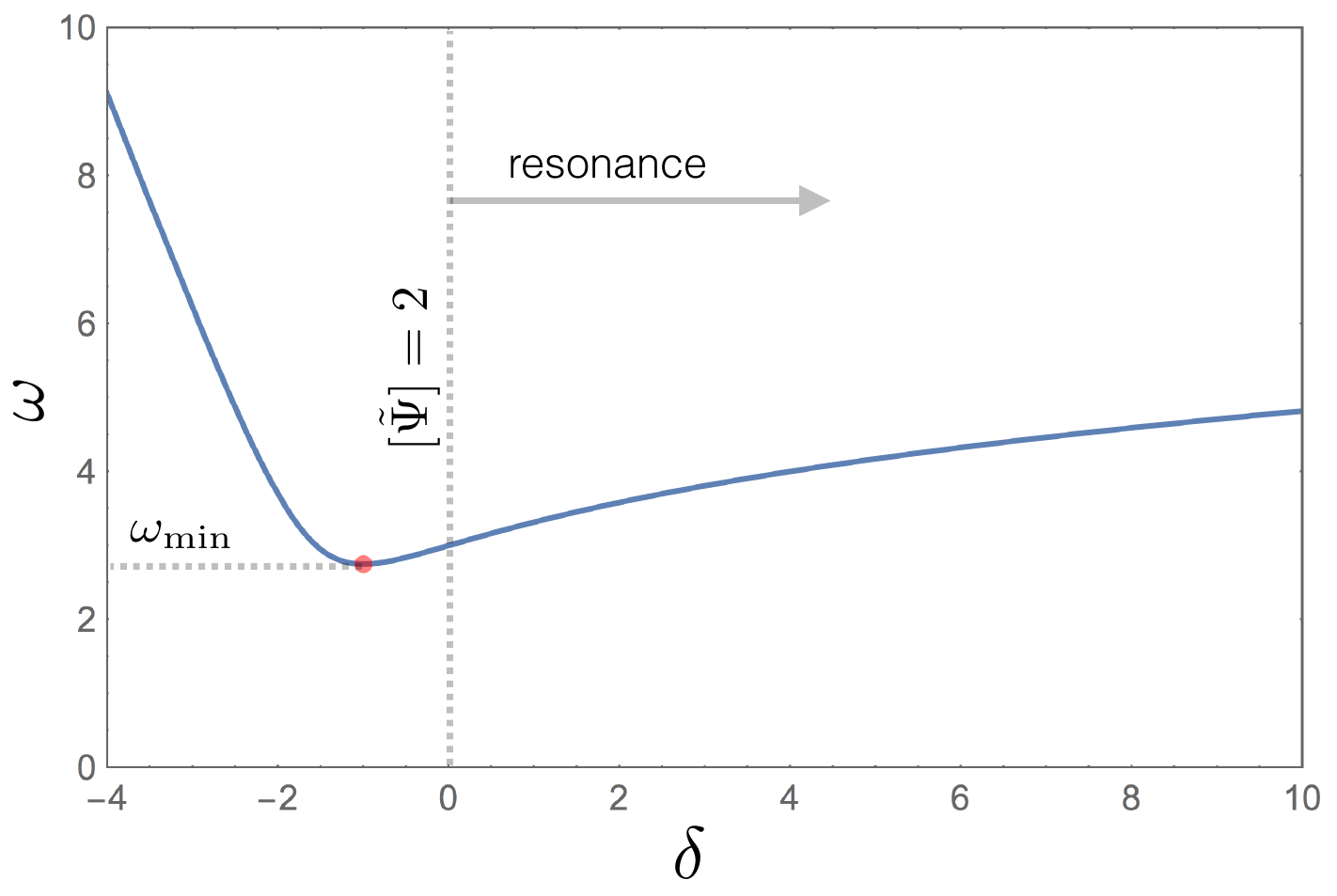}
\caption{The dimensionless libration frequency associated with the resonant equilibrium point. The frequency is minimized at a value of the proximity parameter slightly below $\delta_{\rm{C}}$ and increases slowly and monotonically for $\delta>\delta_{\rm{C}}$. In this work, we adopt the value of $\omega$ at $\delta = \delta_{\rm{C}}$ as the characteristic value.}
\label{freq}
\end{figure}

\subsection{Conditions for Guaranteed Capture}

As already stated above, within the context of Hamiltonian (\ref{SFMR}), the slow convergence of two initially non-resonant orbits towards a mean motion commensurability can be envisioned as a gradual increase in $\delta$ from values below $\delta_{\rm{C}}$ to values above $\delta_{\rm{C}}$ \citep{Peale1986}. Provided that dissipative processes responsible for secular changes in planetary architecture act on much longer timescales than the dynamical timescales associated with resonant motion, we can define an adiabatic invariant \citep{Henrard1982,Neishtadt1984}
\begin{equation}
\mathcal{J} = \oint \PSI\, d\psi = \oint p\, dq,
\end{equation}
which is conserved as long as the trajectory does not encounter a homoclinic curve \citep{Lichtenberg83}. Physically, $\mathcal{J}$ corresponds to the area occupied by the orbit in phase space. 

For a given value of $\delta$, the conservation of  $\mathcal{J}$ dictates the energy level on which the trajectory resides. This is demonstrated in Figure (\ref{portraits}) where a set of orbits characterized by the same $\mathcal{J}$ are shown on phase-space portraits with different values of $\delta$. The condition for guaranteed capture into resonance stems directly from the fact that the adiabatic invariant is preserved as long as the orbit does not encounter a separatrix.  

Of the three thickened circulating trajectories shown in panel A of Figure (\ref{portraits}), consider the one with the smallest radius (i.e. the orbit whose engulfed area is shaded blue and labeled $\mathcal{J}$). This orbit is bound to become a resonant trajectory, (specifically the one corresponding to $\mathcal{J}$ in panels B and C), because the phase-space area it occupies is smaller than the area occupied by the separatrix at $\delta = \delta_{\rm{C}}$, (we shall call this area $\mathcal{J}_{\rm{C}}$), and the total phase-space area occupied by the resonance is minimized when the separatrix first appears \citep{Henrard1982,Morbidelli2002}. Furthermore, because the area occupied by resonant trajectories only grows with increasing $\delta$, the libration amplitude of the trajectory in question shall only decrease with time \citep{Yoder1979,Peale1986}. For reference, $\mathcal{J}_{\rm{C}}$ is shown in green shade in Figure (\ref{portraits}B). 

Let us calculate $\mathcal{J}_{\rm{C}}$. Setting $\delta = \delta_{\rm{C}}$, from equation (\ref{past}) we obtain $\past = 1$. Accordingly, the value of the Hamiltonian that corresponds to the separatrix, $\mathcal{H}_{\rm{C}}^*$, is given by
\begin{equation}
\label{HamC}
\Ham_{\rm{C}}^* = 3(\delta_{\rm{C}}+1) \frac{\left(\past\right)^2}{2}  -\frac{\left(\past\right)^4}{4}- 2 \past = -\frac{3}{4}.
\end{equation}
With a value of $\mathcal{H}_{\rm{C}}$ in hand, the relationship between $\PSI$ and $\psi$ along the separatrix at $\delta = \delta_{\rm{C}}$ can be obtained\footnote{Alternatively, an equivalent relationship between $p$ and $q$ can be obtained from equation (\ref{H}).} from equation (\ref{SFMR}).

\begin{figure*}
\includegraphics[width=1\textwidth]{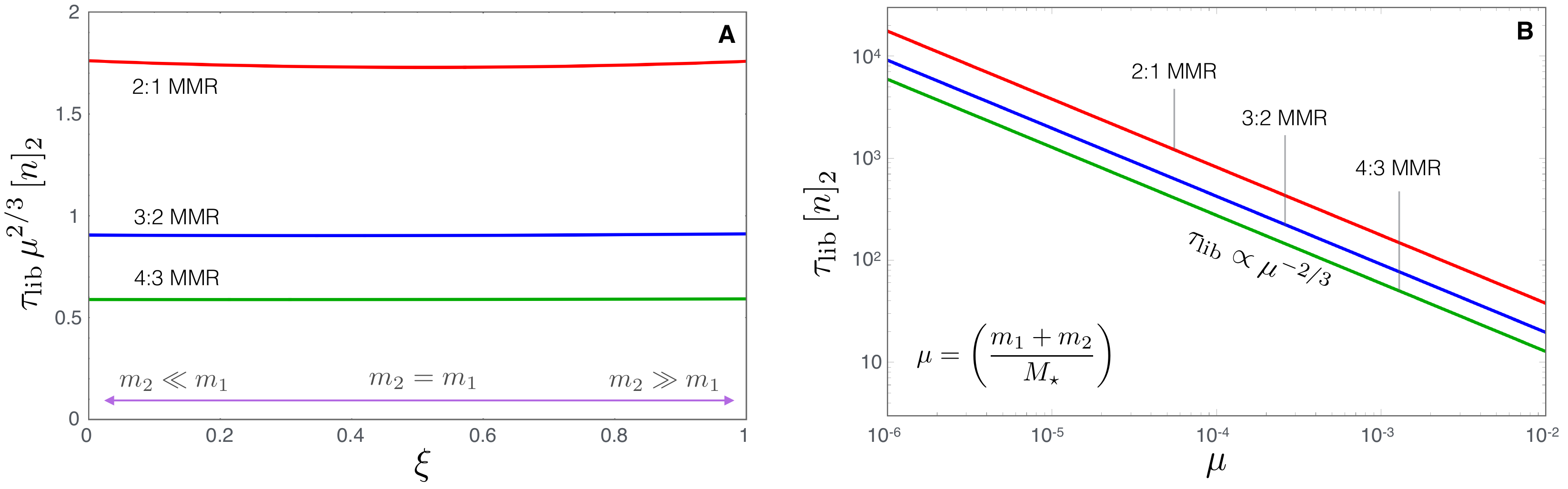}
\caption{Resonant libration period as a function of planet-planet and secondary-primary mass ratios. Panel A shows the dependence of the libration period on how mass is distributed among the two secondary bodies. Panel B depicts the libration period as a function of the total secondary mass to the central mass. Clearly, $\tau_{\rm{lib}}$ is independent of the planet-planet mass ratio to a good approximation. Relationships corresponding to the 2:1, 3:2 and 4:3 resonances are shown as red, blue, and green lines respectively.}
\label{taulib}
\end{figure*}

Integration along the homoclinic curve shows that the phase-space area occupied by the separatrix is given by
\begin{equation}
\mathcal{J}_{\rm{C}} = 6 \left[ \arcsin \left(\frac{1}{(\past)^{3/2}} \right) + \frac{\pi}{2} \right] = 6 \pi.
\end{equation} 
The same procedure yields the area occupied by any trajectory (identified by the value of $\mathcal{H}$) for any value of $\delta$. The evaluation of $\mathcal{J}$ is particularly trivial in the limit where $\delta \ll \delta_{\rm{C}}$. In this case, the phase-space portrait is composed of concentric circles, centered on the stable fixed point near the origin. Therefore, provided some value of $\PSI = p^2/2$, we have 
\begin{equation}
\label{J}
\mathcal{J} = \pi \, p^2 = 2 \pi \, \PSI.
\end{equation}
Naturally, this limit is quite relevant for convergently migrating planets or satellites, as it represents the dynamical state of the system far away from mean motion commensurability. 

The above discussion indicates that the evaluation of $\PSI$ far away from resonance is sufficient to gauge whether or not guaranteed resonant capture will take place. The calculation of $\PSI$ itself requires the specification of planetary eccentricities, longitudes of perihelia as well as planet-star mass ratios. Provided the appropriate quantities, we can write the expression for $\PSI$ in terms of orbital elements and thus formulate a practically useful criterion for guaranteed adiabatic capture within the framework of the unrestricted three-body problem. Specifically, we have:
\begin{empheq}[box=\fbox]{align}
\label{criterion}
&\left[ \frac{3 \, \zeta \M \big( 2 (k-1)^2 m_2 + k^2 m_1 \zeta \big)}{\big(\big(f_{\rm{res}}^{(1)}\big)^2 m_2 + \zeta \big(f_{\rm{res}}^{(2)}\big)^2 m_1\big)^2} \right]^{2/3} \nonumber \\
&\ \ \times\big( \big(f_{\rm{res}}^{(1)} \big)^2 e_1^2 + \big(f_{\rm{res}}^{(2)}\big)^2 e_2^2 \nonumber \\
&\ \ + 2 f_{\rm{res}}^{(1)} f_{\rm{res}}^{(2)} e_1 e_2 \cos(\Delta \varpi) \big)  \leqslant 6,
\end{empheq}
where $\zeta = ((k-1)/k)^{1/3}$. 

The quoted criterion requires knowledge of the planetary eccentricities as well longitudes of perihelia, which is rather inconvenient since these quantities generally depend on the secular dynamics of the planets outside of the resonant domain. Moreover, neither $\PSI$ nor $\Phi_2$ (identified in equation \ref{suicide}) are individually conserved in the secular domain, and large scale variations of the two quantities are indeed possible in principle (see \citealt{BatyginMorbidelli2013AA}). On the other hand, the angular momentum deficit
\begin{equation}
\label{AMD}
\mathcal{A} = \eta\, \PSI + \Phi_2 = m_1 \sqrt{\mathcal{G} \M [a]_1} \, \frac{e_1^2}{2} + m_2 \sqrt{\mathcal{G} \M [a]_2} \, \frac{e_2^2}{2},
\end{equation}
which is more readily interpreted, is conserved on the secular domain. 

Because $\Phi_2 \geqslant 0$, we have: $\PSI \leqslant \mathcal{A}/\eta$. Taking advantage of this, we can formulate an excessively strong, but very simple version of the criterion for guaranteed capture:
\begin{equation}
\label{AMDcrit}
\frac{\mathcal{A}}{\eta} \leqslant 3.
\end{equation}
It is important to note that the maximal extent of orbital excitation that allows for guaranteed capture is a function of the secondary to primary mass ratios. In particular, equations (\ref{criterion}) and (\ref{AMDcrit}) imply that the maximum allowed eccentricities decrease with decreasing mass-ratio, a relationship that may be anticipated intuitively. 

\subsubsection{Circular Restricted Problem as a Special Case} 
Although expression (\ref{criterion}) is derived here for the first time, a related criterion for guaranteed adiabatic capture is well known within the framework of the circular restricted three-body problem (see \citealt{MD99}). Provided that equation (\ref{criterion}) is a more general criterion, it should reduce to the restricted limit as one of the planetary masses is taken to vanish, while the orbit of the other planet is assumed to approach a circle. 

Considering interior resonances first, we set $m_1 = 0$ and $e_2 = 0$ in equation (\ref{criterion}) to obtain: 
\begin{align}
\label{criterion2}
e_1 \leqslant \sqrt{6}\left[\frac{m_2}{3\M}\frac{f_{\rm{res}}^{(1)}}{k^{2/3}(k-1)^{4/3}} \right]^{1/3}.
\end{align}
Similarly, for the case of exterior resonances we set $m_2 = 0$ and $e_1 = 0$, which yields
\begin{align}
\label{criterion3}
e_2 \leqslant \sqrt{6} \left[\frac{m_1}{3 \M}\frac{f_{\rm{res}}^{(2)}}{k^{2}} \right]^{1/3}.
\end{align}
Both of these expressions are in direct agreement with those derived by assuming a circular restricted formalism from the beginning \citep{Peale1986, Malhotra1988}.

It is noteworthy that within the framework of the restricted problem, the aforementioned capture criteria for interior and exterior resonances are obtained from different Hamiltonians. Specifically, the analysis typically begins with a variant of equation (\ref{Horbel}) where only a single resonant harmonic is identified. Subsequently, the corresponding expressions are individually cast into the form of Hamiltonian (\ref{SFMR}) and a calculation of capture probabilities is carried through. In this work, we have utilized the reducing transformation (\ref{suicide}) to derive the previously known criteria as limiting cases of a more comprehensive formalism. Our analysis thus generalizes previous results.

\subsection{Probabilistic Capture}

If the criterion (\ref{criterion}) is not satisfied, resonant capture is not certain. This is because at sufficiently high values of the action $\PSI$, the phase-space area occupied by the orbit exceeds the critical area occupied by the separatrix at the inception of the resonance i.e. $\mathcal{J} > \mathcal{J}_{\rm{C}}$ at $\delta = \delta_{\rm{C}}$. Consequently, unlike the ``smooth" transition to resonance discussed above, in this case the trajectory must cross the separatrix in order to reach the resonant domain. However, passage across a critical curve is a fundamentally probabilistic process, and may advect the trajectory into the inner circulation region of the phase-space portrait.

To determine the probabilistic outcome of an encounter of a $\mathcal{J} > \mathcal{J}_{\rm{C}}$ trajectory with a homoclinic curve, one must first relate the non-resonant initial conditions to the value of the proximity parameter at which the transition will take place. Utilizing the conservation of the adiabatic invariant (equation \ref{J}), we can connect the value of the action $\PSI_0$ far away from resonance (i.e. at $\delta \ll \delta_{\rm{C}}$, where the trajectory in phase-space can be approximated by a circle) to the value of the proximity parameter at which the encounter will take place $\delta_{\rm{enc}}$. This can be done by matching the phase space area occupied by the trajectory to that engulfed by the separatrix at $\delta_{\rm{enc}}$. The appropriate expression reads \citep{Henrard1993}:
\begin{align}
\label{areamatch}
&\PSI_0 = \frac{1}{2} \left[ \frac{3 \, \zeta \M \big( 2 (k-1)^2 m_2 + k^2 m_1 \zeta \big)}{\big(\big(f_{\rm{res}}^{(1)}\big)^2 m_2 + \zeta \big(f_{\rm{res}}^{(2)}\big)^2 m_1\big)^2} \right]^{2/3} \nonumber \\
&\times \bigg( \big(f_{\rm{res}}^{(1)} \big)^2 e_1^2 + \big(f_{\rm{res}}^{(2)}\big)^2 e_2^2  + 2 f_{\rm{res}}^{(1)} f_{\rm{res}}^{(2)} e_1 e_2 \cos(\Delta \varpi) \bigg)  \nonumber \\
&= \frac{3}{\pi}\bigg((\delta_{\rm{enc}}+1)\left[\arcsin \left(\frac{1}{(\past)^{3/2}} \right) + \frac{\pi}{2} \right] \nonumber \\
&\ \ \ + \frac{\sqrt{(\past)^3-1}}{\past} \bigg).
\end{align}
Recalling that $\past$ is a single-valued function of $\delta$ (see equation \ref{past}), the above expression unequivocally links the starting condition of the system to the phase-space portrait at which the encounter takes place. Figure (\ref{pcapt}A) shows $\mathcal{J}/2\pi=\PSI_0$ as a function of $\delta_{\rm{enc}}$, obtained from equation (\ref{areamatch}).

With the above relationship at hand, we can evaluate the probability of resonant capture for arbitrary initial conditions. Generally, adiabatic capture probability, $\mathcal{P}$, is given by the ratio of the rates of changes of the phase-space areas occupied by the resonant domain of the phase-space portrait, to that occupied by the outer branch of the separatrix (see Figure \ref{portraits}C). Following \citet{BorderiesGoldreich1984}, the explicit expression\footnote{It should be understood that within the framework of this expression, a value of $\mathcal{P}$ that exceeds unity simply corresponds to certain capture.} for $\mathcal{P}$ that corresponds to the Hamiltonian (\ref{SFMR}) reads:
\begin{empheq}[box=\fbox]{align}
\label{Pcapt}
\mathcal{P} = 2 \left[1+\frac{\pi}{2 \arcsin\big((\past)^{-3/2} \big)} \right]^{-1}.
\end{empheq}

Figure (\ref{pcapt}B) depicts the resonant capture probability as a function of $\PSI_0$. It should be noted that in the case of probabilistic capture, the transition through the separatrix necessarily breaks the conservation of $\mathcal{J}$. This means that the resonant encounter will lead to an impulsive change in orbital parameters, irrespective of whether capture is successful. To this end, note that passage through the resonance (see Figure \ref{portraits}C) diminishes the overall phase-space area, implying a less dynamically excited post-encounter state.

\begin{figure*}
\includegraphics[width=1\textwidth]{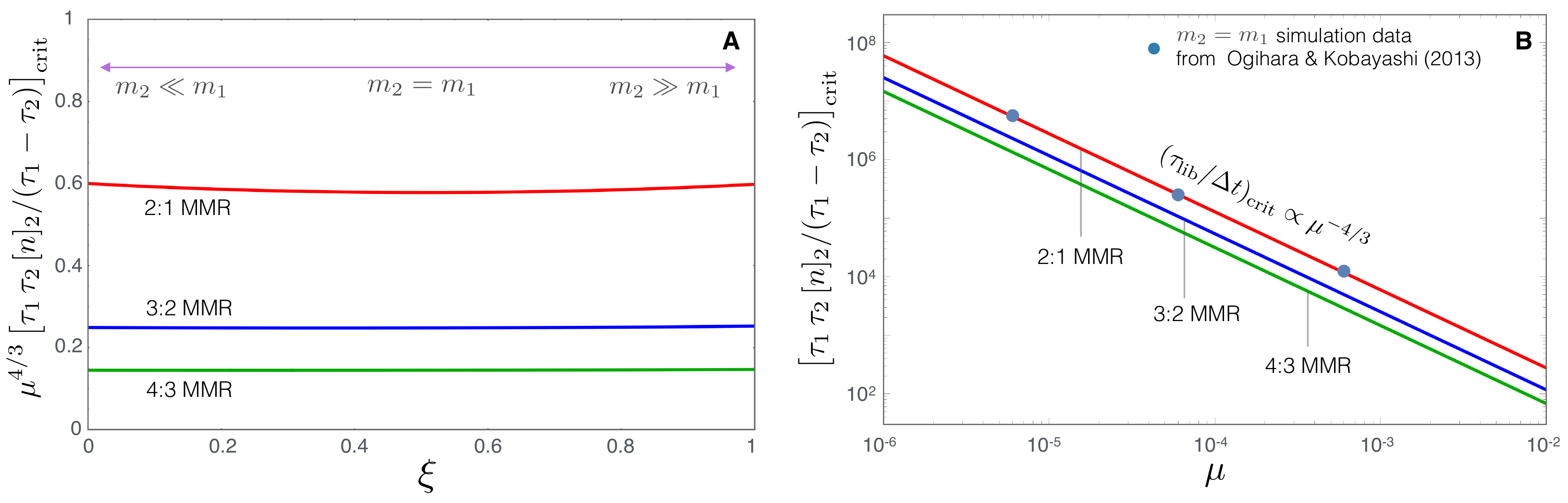}
\caption{The adiabatic threshold as a function of plane-planet and planet-star mass ratios. As in Figure (\ref{taulib}), panel A depicts the dependence of the adiabatic criterion on the distribution of mass among the planets, while panel B shows its $-4/3$ power-law dependence on the ratio of the total planet mass to the central mass. In addition to the theoretical relationships, numerically computed adiabatic threshold data corresponding to equal-mass bodies in a 2:1 resonance \citep{OgiharaKobayashi2013} are shown with blue points. Curves corresponding to the 2:1, 3:2 and 4:3 resonances are shown as red, blue, and green lines respectively.}
\label{ad_crit}
\end{figure*}

\subsection{The Adiabatic Threshold}
The entirety of the above discussion assumes that the dissipative evolution of the proximity parameter is so slow that at any point, a ``frozen" (i.e. constant $\delta$) representation of the dynamics provides a good approximation to the real evolution \citep{Wisdom1985,Henrard1990}. This approximation holds in a regime where the resonant libration frequency greatly exceeds the extrinsic resonance crossing rate. Within the framework of the adopted model, both quantities can be analytically derived, thereby providing an explicit criterion for adiabatic dynamics.

To calculate the resonant libration frequency, we expand the Hamiltonian (\ref{SFMR}) to second order in $\PSI$ and $\psi$ around the equilibrium point $([\PSI],\pi)$ that corresponds to the definite negative root of equation (\ref{equilibrium}). Dropping constant terms, we have: 
\begin{align}
\Ham &= -\left(1+ \frac{1}{2\sqrt{2}[\PSI]^{3/2}} \right)(\PSI-[\PSI])^2 - \sqrt{2 [\PSI]} (\psi-\pi)^2 \nonumber \\
&+\bigg(3+ 3 \delta -2 [\PSI] + \sqrt{\frac{2}{[\PSI]}} \bigg) (\PSI-[\PSI]).  \\
\nonumber
\end{align}
The fixed-point condition dictates that in the above expression, the term linear in $\PSI$ must vanish. This yields a relationship between the proximity parameter and the equilibrium action:
\begin{align}
\label{vanish}
\delta &= \frac{2}{3}[\PSI] - \frac{1}{3}\sqrt{\frac{2}{[\PSI]}} -1.
\end{align}

Introducing new canonical variables (e.g. \citealt{BatyginMorbidelli2013AA}):
\begin{align}
\bar{\Psi} &= (\PSI - [\PSI]) \left( \frac{\sqrt{2} + 4 [\PSI]^{3/2}}{4 \sqrt{2} [\PSI]^2} \right)^{1/4} \nonumber \\
\bar{\psi} &= (\psi - \pi) \left( \frac{\sqrt{2} + 4 [\PSI]^{3/2}}{4 \sqrt{2} [\PSI]^2} \right)^{-1/4},
\end{align}
we obtain a Hamiltonian that is synonymous to that of a simple harmonic oscillator: 
\begin{align}
\Ham &= - \frac{\omega}{2} \left( \bar{\Psi}^2 + \bar{\psi}^2 \right),
\end{align}
where 
\begin{align}
\omega &= \sqrt{\frac{2+4 \sqrt{2} [\PSI]^{3/2}}{[\PSI]}}
\end{align}
is identified as the resonant libration frequency \citep{Lichtenberg83}. 

Utilizing equation (\ref{vanish}), $\omega$ can be expressed as a function of the proximity parameter, $\delta$, and this relationship is shown in Figure (\ref{freq}). It is trivial to show that the value of $\omega$ is minimized at $\delta = -1$ (corresponding to $[\PSI]=2^{-1/3}$) and quantitatively evaluates to $\omega_{\rm{min}} = 2^{2/3}\sqrt{3}$. For larger values of $\delta$, $\omega$ slowly increases. For example, at the onset of resonance (i.e. at $\delta = 0$ when the separatrix first appears), $\omega = 3$. Recalling the scalings introduced in equations (\ref{scale}) and (\ref{eta}), we adopt the value of the libration frequency at $\delta_{\rm{C}}$ to formulate a practically useful criterion for adiabatic evolution\footnote{Non-adiabatic evolution will cause the trajectory to jump to the inner circulation region without being trapped into resonance. So, it is sensible to consider the libration frequency at a value of $\delta$ at which the inner circulation region first appears i.e. $\delta_{\rm{C}}$. At this value of $\delta$, $\omega$ exceeds its minimum value by $\sim 10\%$.}. Specifically, in terms of physical parameters, the corresponding resonant libration period is given by:
\begin{align}
\tau_{\rm{lib}} &= \frac{4 \pi}{3 \, [n]_2} \frac{(k-1)^{2/9}}{3^{1/3} k^{5/9}} \nonumber \\
&\times \bigg( \frac{\M}{((k-1)/k)^{1/3} \big(f_{\rm{res}}^{(2)}\big)^2 m_1 + \big(f_{\rm{res}}^{(1)}\big)^2 m_2}  \bigg)^{1/3} \nonumber \\
&\times \bigg( \frac{\M}{((k-1)k^2)^{1/3} m_1 + (k-1) m_2}  \bigg)^{1/3}.
\end{align}

It is instructive to examine the dependence of $\tau_{\rm{lib}}$ on the distribution of mass between $m_1$ and $m_2$. The relationship can be evaluated by setting $m_1 = \mu \, (1- \xi) \, \M$ and $m_2 = \mu\, \xi \, \M$, where $\mu = (m_1 + m_2)/\M$, and examining the dependence of $\tau_{\rm{lib}}$ on $\xi$. Figure (\ref{taulib}A) shows $\tau_{\rm{lib}}$, appropriately scaled by the mass ratio parameter and the outer planet's orbital period for 2:1, 3:2, and 4:3 resonances. As can be immediately gleaned, the libration period is independent of $\xi$ to an excellent approximation\footnote{Specifically, the fractional variation is of order $\sim 1\%$.}. Thus, we can simplify the above expression somewhat by replacing $m_2$ with the cumulative mass of the planets and setting $m_1 = 0$:
\begin{align}
\label{taulibapprox}
\tau_{\rm{lib}} &\simeq \frac{4 \pi}{3 \, [n]_2}\left(\frac{\M}{m_1 + m_2} \right)^{2/3} \left[\frac{(3 \big(f_{\rm{res}}^{(1)}\big)^2)^{-1/3}}{(k^5 (k-1))^{1/9}}\right].
\end{align}
Figure (\ref{taulib}B) shows the resonant libration period as a function of the mass-ratio.

In order to obtain a condition for adiabatic evolution, we require an additional piece of information: the characteristic timescale on which dissipative forces will carry the system across the resonance width, in absence of planet-planet interactions \citep{Friedland2001}. As a first step, we must define the resonance width, $\Delta \Psi$. The maximal excursion in $\PSI$ at $\delta = 0$ corresponds to the difference between the points at which the critical curve crosses the $p-$axis. Noting that one of the crossing points is the bifurcated equilibrium, from equations (\ref{past}) and (\ref{HamC}), we have:
\begin{align}
\label{reswidthyo}
\Delta \Psi = \left(\frac{(p_{*})^2}{2} - \frac{(\past)^2}{2} \right) \eta = 4 \, \eta,
\end{align}
where $p_{*} = - 3$ signifies the negative intersection point of the separatrix. 

From conservation of the integrals $\mathcal{K}$ and $\Omega$ (see equations \ref{transformone} and \ref{finalactionangle}), it follows that the maximal resonant variation of the first \Poincare\ momenta is
\begin{align}
\Delta \Lambda_1 &=  - (k-1) \, \Delta \Psi \, \eta \nonumber \\
\Delta \Lambda_2 &=   k \, \Delta \Psi \, \eta.
\end{align}
The quantity of interest for our calculation is the ratio of mean-motions:
\begin{align}
\chi &= \frac{n_2}{n_1} \simeq \frac{[n]_2}{[n]_1}\bigg(1 + 3  \frac{(\Lambda_1 - [\Lambda]_1)}{[\Lambda]_1} - 3  \frac{(\Lambda_2 - [\Lambda]_2)}{[\Lambda]_2} \bigg).
\end{align}
Accordingly, the above expressions define a resonance width in $\chi$:
\begin{align}
\left(\Delta \chi \right)_{\rm{res}} &=  \frac{[n]_2}{[n]_1}\bigg(1 - 3 (k-1) \frac{\Delta \Psi \eta}{[\Lambda]_1} - 3 k \frac{\Delta \Psi \eta}{[\Lambda]_2} \bigg) \nonumber \\
&=  - 12 \eta \, \frac{k \, ([\Lambda]_1+[\Lambda]_2) - [\Lambda]_2}{[\Lambda]_1 \, [\Lambda]_2}.
\end{align}

Let us now turn to dissipative evolution. A conventional way to parameterize extrinsically facilitated drifts in semi-major axes is to adopt the following form \citep{LeePeale2002}: 
\begin{align}
\frac{da_1}{dt} = \frac{a_1}{\tau_1} \ \ \ \ \ \ \ \ \ \  \frac{da_2}{dt} = \frac{a_2}{\tau_2}
\end{align}
From these expressions, it follows that the associated rate of change in the ratio of mean motions is:
\begin{align}
\frac{d \chi}{dt} = - \frac{3}{2} \frac{(\tau_1 - \tau_2)}{ \tau_1 \tau_2} \frac{n_2}{n_1}.
\end{align}
By replacing the derivative on the LHS by a fraction of finite differences, and equating the change in $\chi$ to the resonance width, we obtain a corresponding segment of time required for dissipative forces to carry the orbits across the resonance \citep{Friedland2001,GoldreichSchlichting2014}: 
\begin{align}
\Delta t \simeq 8 \eta\, \frac{(\tau_1 - \tau_2)}{ \tau_1 \tau_2} \frac{k \, ([\Lambda]_1+[\Lambda]_2) - [\Lambda]_2}{[\Lambda]_1 \, [\Lambda]_2} .
\end{align}

The condition for adiabatic evolution is thus defined: the characteristic timescale of orbital convergence, $\Delta t$, must exceed the libration period $\tau_{\rm{lib}}$. In terms of physical parameters, the adiabatic criterion reads:
\begin{empheq}[box=\fbox]{align}
\label{adiabaticcrit}
&\frac{\tau_{\rm{lib}}}{\Delta t} = \frac{(k-1)^{4/9} \, \pi}{ 2 \, (3)^{2/3} \, [n]_2} \frac{(\tau_1 - \tau_2)}{ \tau_1 \tau_2} \nonumber \\
&\times \bigg( \frac{\M}{((k-1)/k)^{1/3} \big(f_{\rm{res}}^{(2)}\big)^2 m_1 + \big(f_{\rm{res}}^{(1)}\big)^2 m_2}  \bigg)^{2/3} \nonumber \\
&\times \bigg( \frac{\M}{((k-1)k^2)^{1/3} m_1 + (k-1) m_2}  \bigg)^{2/3} \lesssim 1.
\end{empheq}

Similarly to the preceding discussion of the expression for $\tau_{\rm{lib}}$, equation (\ref{adiabaticcrit}) is approximately independent of how mass is distributed between the secondary bodies. Instead, the criterion depends sensitively on the ratio of the total planetary mass to the mass of the central object. Accordingly, in the same spirit as equation (\ref{taulibapprox}), we can obtain a simplified expression for the adiabatic criterion:
\begin{align}
\label{adiabatapprox}
\frac{\tau_{\rm{lib}}}{\Delta t} &\simeq \frac{\pi}{2\, [n]_2} \frac{(\tau_1 - \tau_2)}{ \tau_1 \tau_2} \left(\frac{\M}{m_1 + m_2} \right)^{4/3}  \nonumber \\
&\times \left[ \frac{(k-1)^{-2/9}}{\big(\sqrt{3} f_{\rm{res}}^{(1)} \big)^{4/3} } \right] \lesssim 1.
\end{align}
We note however, that dependence on the planet-planet mass ratio can still enter equations (\ref{adiabaticcrit}) and (\ref{adiabatapprox}) through particular expressions for $\tau_1$ and $\tau_2$.

\begin{figure}
\includegraphics[width=1\columnwidth]{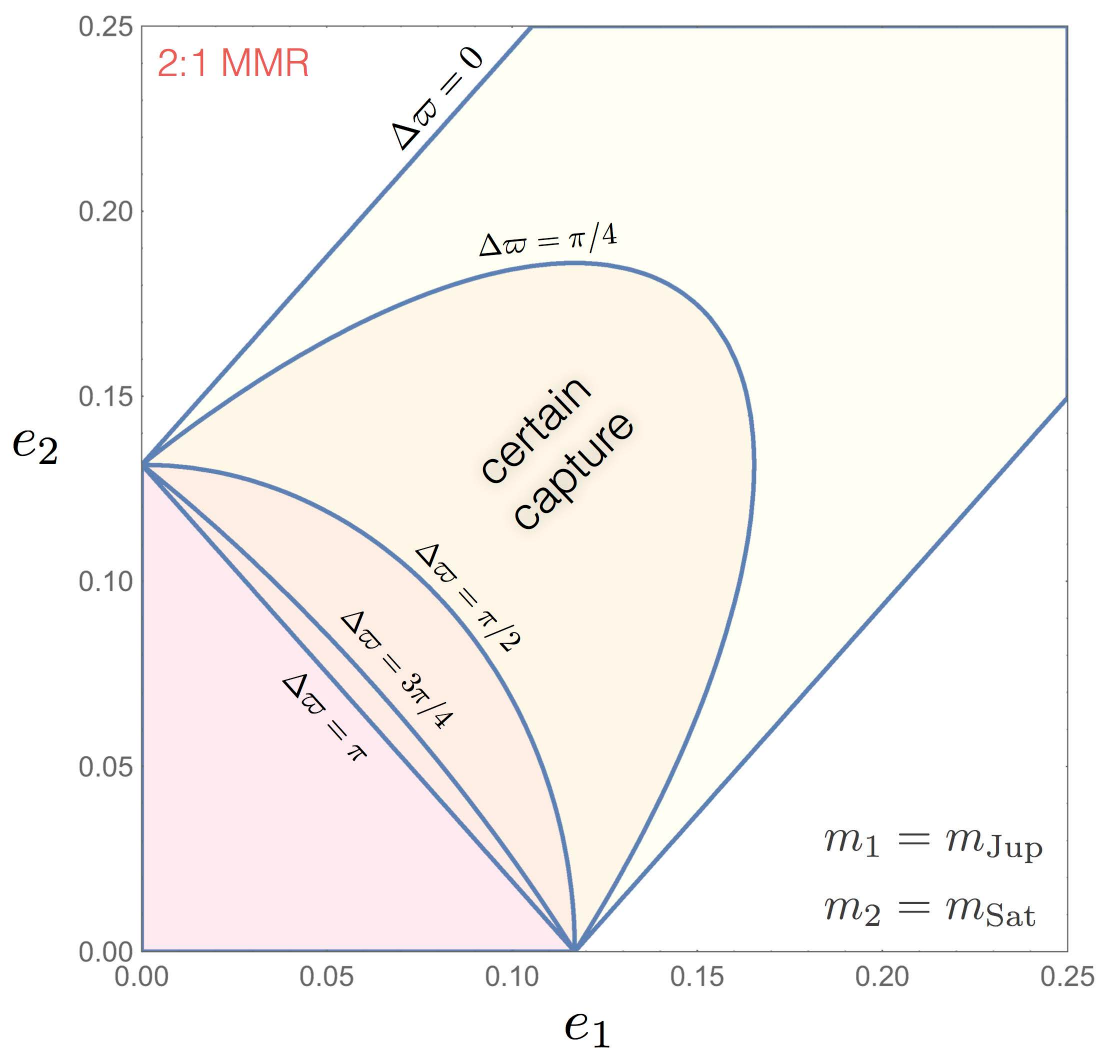}
\caption{Conditions for guaranteed adiabatic capture of Jupiter and Saturn into the 2:1 mean motion resonance. Each of the plotted curves corresponds to a distinct assumed difference in the apsidal lines, $\Delta \varpi$. The $e_1-e_2$ parameter space associated with certain capture is clearly minimized at $\Delta \varpi= \pi$ and maximized at $\Delta \varpi= 0$.}
\label{JS_21fig}
\end{figure}

The analytical expression for the adiabatic threshold derived above exhibits the characteristic $-4/3$ power-law dependence on the mass ratio. Unsurprisingly, the same power-law holds for the restricted problem \citep{Friedland2001,Quillen2006,GoldreichSchlichting2014}, and has been observed to arise in numerical N-body simulations \citep{Ketchum2011,OgiharaKobayashi2013}. Figure (\ref{ad_crit}A) shows the dependence of the adiabatic criterion on the distribution of masses, while the dependence on the cumulative secondary to primary mass ratio is shown in Figure (\ref{ad_crit}B). Also over-plotted on panel B are numerical estimates of the adiabatic threshold for two equal-mass bodies corresponding to the 2:1 resonance obtained by \citet{OgiharaKobayashi2013}. Clearly, the agreement between theory and simulation is more than satisfactory. 

\section{Results}

With the crucial features of the analytical theory defined, we are now in a position to examine the generic consequences of the model. Additionally, we shall aim to understand how observations of exoplanetary systems and simulations of the early dynamical evolution of the Solar System fit into the framework of the developed formalism. We shall begin by outlining some generalities.

\subsection{General Results}
\subsubsection{Capture Probability Maps}

The formulae presented in the previous section allow for the evaluation of $\mathcal{P}$ at a computationally negligible cost. Taking advantage of this, we are provided with an opportunity to rapidly delineate the probability of capture in a given section of parameter space, a task that would be impossible to accomplish with brute-force numerical simulations. To this end, the only quantity we need to specify is $\PSI$ far away from resonance. However, $\PSI$ itself represents a rather complicated combination of orbital and physical parameters, and involves the specification of planet-star and planet-planet mass-ratios, orbital eccentricities, as well as the difference in the longitudes of perihelia. 

Although the specification of mass-ratios and eccentricities is unavoidable, one would ideally like to be agnostic with respect to specifying the difference in longitudes of perihelia of the orbits, $\Delta \varpi$. Unfortunately, the outcome of resonant encounters is not insensitive to $\Delta \varpi$. As an example, consider a pair of planets with star-planet mass ratios equal to that of Jupiter and Saturn in the vicinity of a 2:1 commensurability (this example will also be relevant to the discussion presented below). Utilizing equation (\ref{criterion}), we can delineate curves that correspond to guaranteed capture on a $e_1$ vs $e_2$ digram, as shown in Figure (\ref{JS_21fig}). Clearly, parameter space associated with certain capture is minimized for $\Delta \varpi = \pi$ and maximized for $\Delta \varpi = 0$.

\begin{figure*}
\includegraphics[width=1\textwidth]{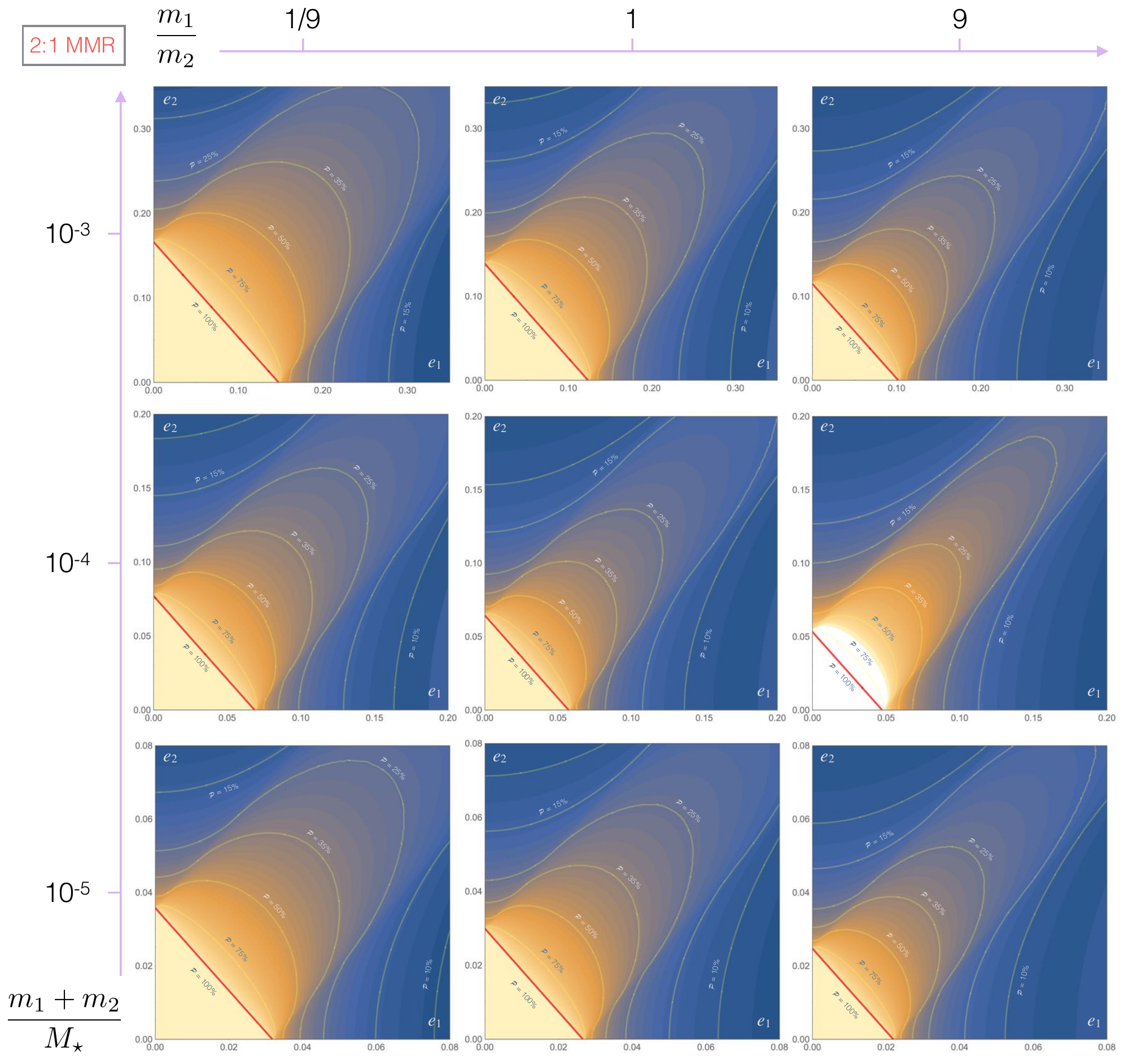}
\caption{A series of maps of the apsidally averaged capture probability corresponding to the 2:1 mean motion resonance, for a variety of planet-planet and planet-star mass ratios. The color scale marks the averaged capture probability, with orange and blue corresponding to high and low values respectively. Additionally, on each plot a $\langle\mathcal{P}\rangle = 100\%$, curve below which capture is certain is shown with a thick red line, while a series of contours corresponding to $\langle\mathcal{P}\rangle = 75\%, 50\%, 35\%,  25\%, 15\%, 10\%$ are shown as gold curves. Note that the range of the plots increases with total secondary to primary mass. Specifically, $e_{\rm{max}} = 0.08, 0.2, 0.35$ for $(m_1+m_2)/\M = 10^{-5}, 10^{-4},10^{-3}$ respectively.}
\label{map21}
\end{figure*}

\begin{figure*}
\includegraphics[width=1\textwidth]{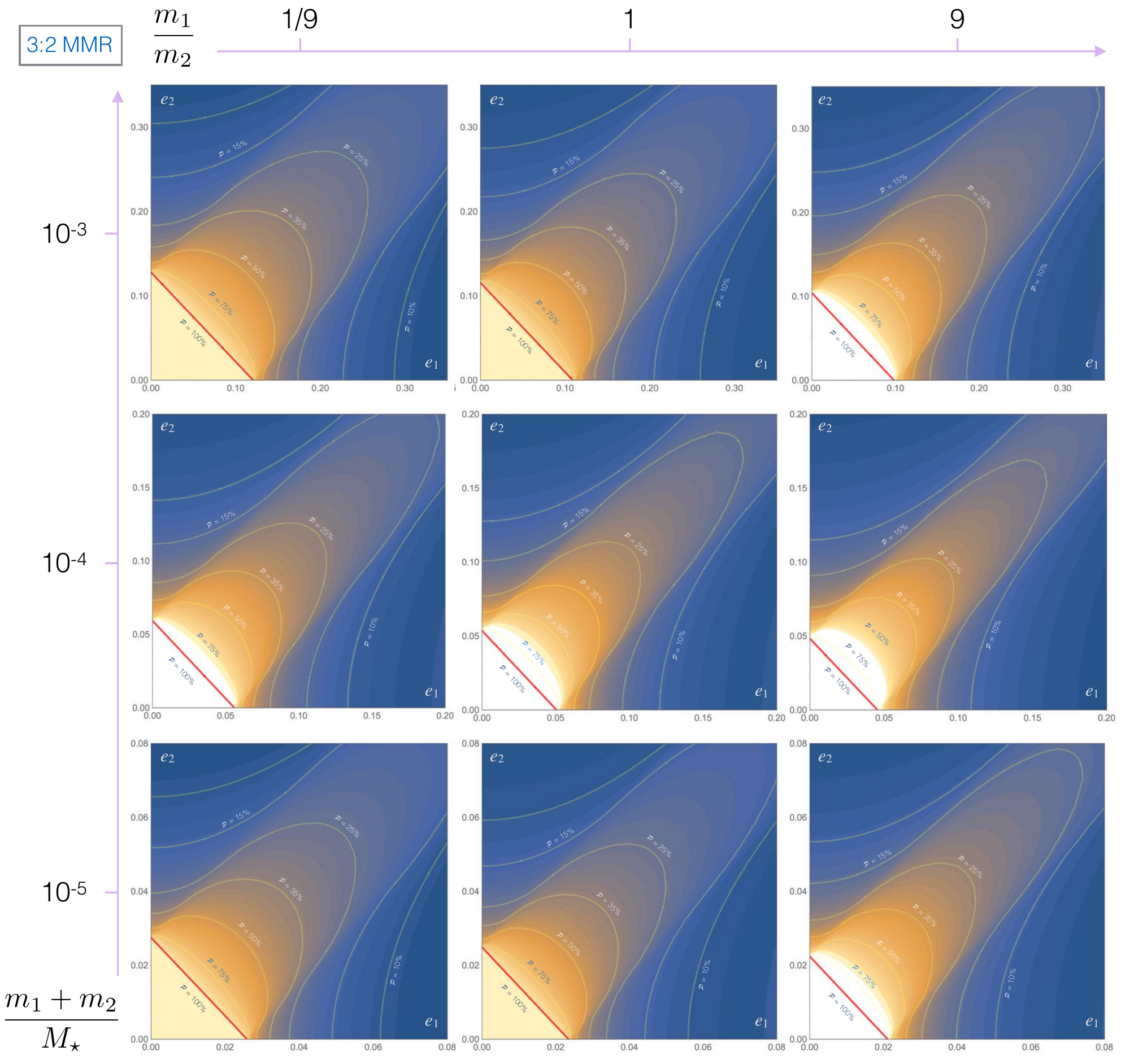}
\caption{Same as Figure (\ref{map21}), but for the 3:2 mean motion resonance.}
\label{map32}
\end{figure*}

\begin{figure*}
\includegraphics[width=1\textwidth]{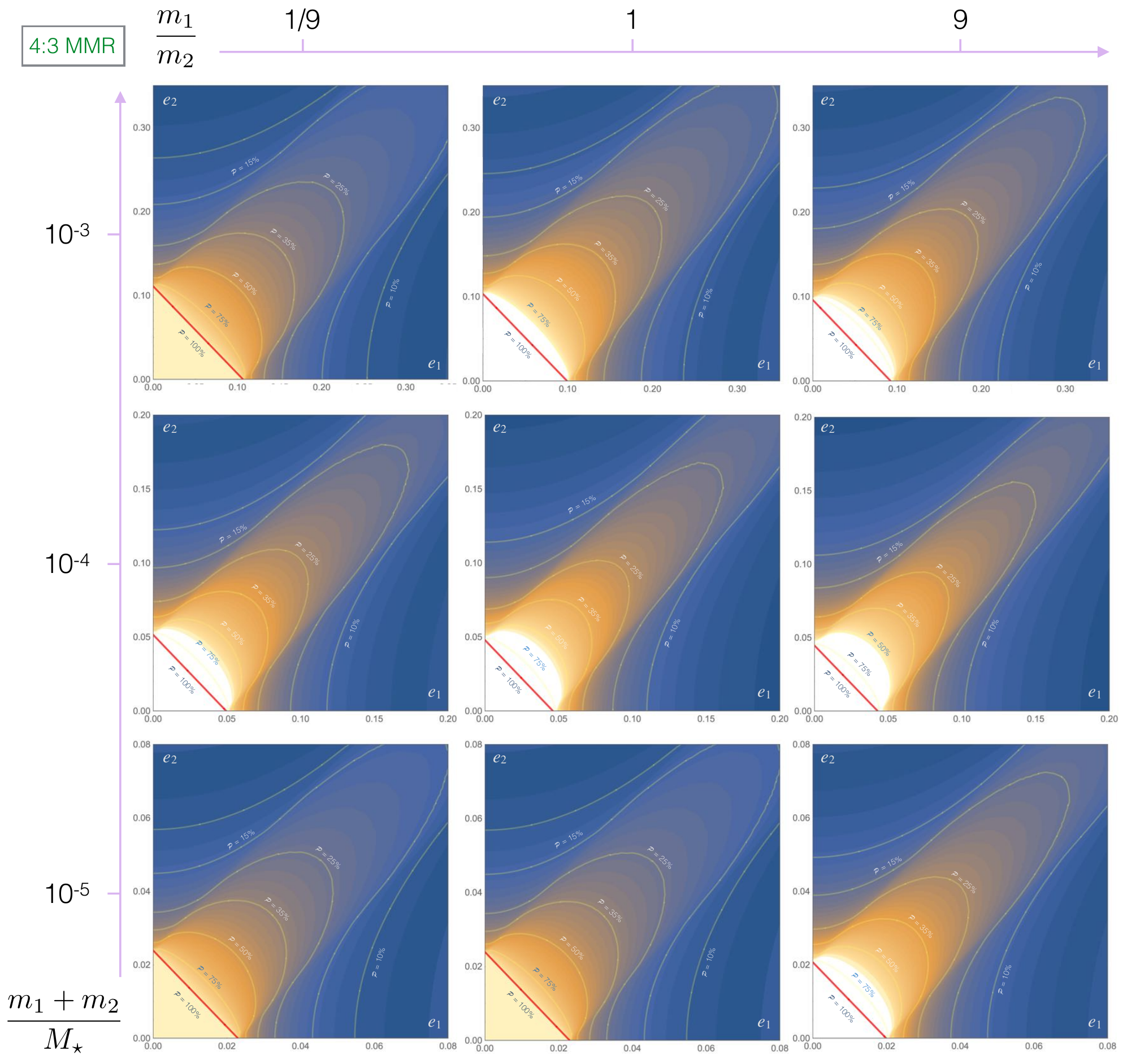}
\caption{Same as Figure (\ref{map21}), but for the 4:3 mean motion resonance.}
\label{map43}
\end{figure*}

Thus, if the knowledge of $\Delta \varpi$ is readily accessible, one may proceed with the estimation of capture probabilities in the conventional fashion. If not however, it is sensible to assume that this difference of angles is uniformly distributed between $0$ and $2\pi$, and correspondingly average over $\Delta \varpi$, leading to the following definition of the apsidally-averaged capture probability:
\begin{align}
\label{Pcaptave}
\langle \mathcal{P} \rangle = \frac{1}{2\pi} \int_{0}^{2\pi} \mathcal{P}\, d \Delta \varpi.
\end{align}
With this definition at hand, the number of input parameters is reduced to four (specifically, they are the total secondary mass $(m_1+m_2)/\M$, the secondary mass-ratio $m_1/m_2$, and the two orbital eccentricities $e_1$ and $e_2$) and can thus be represented on a series of 2D level contour plots.

We have calculated the averaged probability across a range of parameters for the 2:1, 3:2 and the 4:3 mean motion resonances. In particular, we have mapped out $\langle \mathcal{P} \rangle$ on a chain of $e_1$ vs $e_2$ diagrams, corresponding to total secondary masses of $(m_1+m_2)/\M = 10^{-3}, \,10^{-4}, \, 10^{-5}$ and secondary mass-ratios of $m_1/m_2 = 1/9,1,9$. The results are shown in Figures (\ref{map21}) - (\ref{map43}). 

On each diagram, the orange-blue color-scale denotes the averaged capture probability, which is also marked by distinct yellow contour lines. A thickened red line that represents the conditions for guaranteed capture is also shown in each panel. Importantly, within the section of parameter space restricted by the red line, any combination of orbital eccentricities and differences in apsidal lines will necessarily lead to resonant locking. Note also that the range over which $e_1$ and $e_2$ are plotted diminishes with decreasing total secondary mass.

Upon examination of Figures (\ref{map21}) - (\ref{map43}), a few features immediately stand out. First, the overall capture probability decreases with $k$, meaning that capture into the 2:1 mean motion resonance is the most generous with respect to the degree of pre-encounter orbital excitation. A second readily apparent quality is the weak dependence of $\langle \mathcal{P} \rangle$ on the distribution of masses among the two bodies. That is, the difference between the values of eccentricities that characterize guaranteed capture is smaller than a factor of two, depending on whether the majority of the mass resides in the inner or the outer planet. Moreover, the dependence on $m_1/m_2$ further subsides with increasing $k$. Third, it is clear that the probability of capture most strongly depends on the total planetary mass of the system.

The construction of capture probability maps in principle removes the necessity for computationally expensive numerical experiments, provided that the assumptions inherent to the theory are satisfied for the problem at hand. The results also highlight an important attribute of resonant dynamics: while substantial orbital excitation ($e\sim 0.1$) does not inhibit capture for giant planets, rather minute ($e\sim 0.02$) eccentricities may be sufficient to obstruct the onset of resonant evolution of Earth-like terrestrial planets. We shall return to this point again below. 

\subsubsection{Critical Rates of Convergent Evolution}

An implicit assumption inherent to the calculation of capture probability maps depicted in Figures (\ref{map21}) - (\ref{map43}) is that orbital convergence is much slower than angular momentum and energy exchange via resonant interaction. As shown in the previous section, it is possible to write down an analytic criterion for adiabatic evolution. Accordingly, we shall now apply this criterion to the most common mechanisms responsible for orbital migration (namely interactions of planets with gaseous disks, scattering of planetesimals, and tidal evolution), and derive a series of constraints on physical parameters inherent to each physical setting. 

\smallskip
$\bullet$ \textit{Disk-Driven Migration}
\smallskip

Gravitational interactions between a planet and a gaseous disk are well-known to yield a time-irreversible exchange of angular momentum \citep{GoldreichTremaine1980}. At the detailed level, the rate and direction of migration are determined by the physical properties of the disk (e.g. density and entropy profiles) as well as the planetary mass \citep{Paardekooper2010,Paardekooper2011,Bitsch2011}. Depending on the latter, disk-driven orbital migration typically falls into one of two\footnote{For the purposes of this work, we shall neglect the so-called ``type-III" mode of orbital transport.} distinct regimes \citep{Ward1997,PapaloizouLarwood2000}. For planets that are not sufficiently massive to clear out their own orbital neighborhoods and thereby carve out gaps in their gaseous disks, migration proceeds in the so-called ``type-I," or linear regime. We consider this mode of orbital transport first.

A rudimentary way to define a planetary mass below which type-I migration applies is to consider the viscous gap-opening criterion:
\begin{align}
\label{gapopen}
\frac{m}{\M} \lesssim \sqrt{\frac{27 \pi}{8}\left(\frac{h}{a} \right)^5 \bar{\alpha}},
\end{align}
where $h$ is the scale-height of the disk, and $\bar{\alpha}$ is the Shakura-Sunayev viscosity parameter \citep{Armitage2010}. The quoted expression is derived by equating resonant and viscous torques in the planetary neighborhood. A more sophisticated treatment of the gap-opening process can be found in \citep{Crida2006}.

\begin{figure}
\includegraphics[width=1\columnwidth]{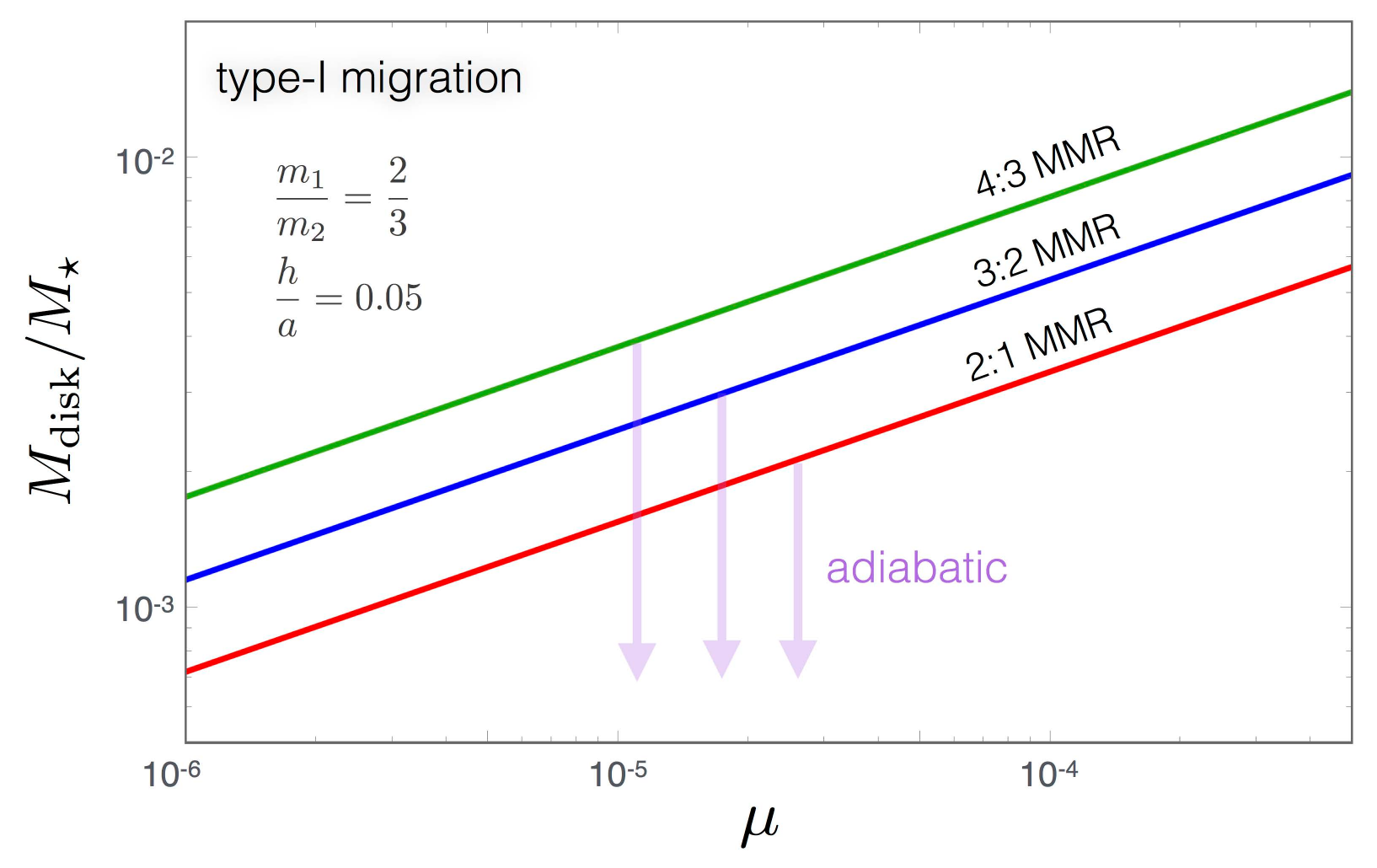}
\caption{The adiabatic threshold for two objects undergoing type-I migration in an isothermal disk with a Mestel-like surface density profile. While this particular figure shows the adiabatic limit for a secondary mass-ratio of $m_1/m_2 = 2/3$ ($\xi = 3/5$), it should be kept in mind that unlike the expression for $\tau_{\rm{lib}}$, equation (\ref{adiabatTypeI}) does depend strongly on this quantity.}
\label{typeIfig}
\end{figure}

Under the simplifying assumptions of a power-law surface density profile and an isothermal equation of state, the direction of type-I migration is strictly inwards, and occurs on a characteristic timescale \citep{Tanaka2002}:
\begin{align}
\label{typeI}
\tau_{\rm{I}} = - \frac{1}{s} \frac{\M^2}{m \, \Sigma \, a^2} \left(\frac{h}{a} \right)^2 \frac{1}{n},
\end{align}
where $s$ is a dimensionless constant that depends on the structure of the disk. 

Motivated by observations of \citet{Andrews2013}, here we shall assume that the surface density profile takes the form of a classical Mestel disk i.e. $\Sigma = \Sigma_0 \, a_0/a$, where $\Sigma_0$ is the surface density at some reference semi-major axis $a_0$ \citep{Mestel1963}. With this choice for $\Sigma$, the have $s = 3.8$ \citep{Tanaka2002}. Additionally, for simplicity we shall assume that the disk aspect ratio, $h/a$, is constant throughout the region of interest. 

From the form of equation (\ref{typeI}), it is immediately clear that if two planets are undergoing type-I orbital decay, a necessary requirement for convergent migration is $m_2 > \zeta \, m_1$. Assuming that this criterion is satisfied, we evaluate the decay timescales (\ref{typeI}) at nominal resonance, such that $a_1 = ((k-1)/k)^{2/3} = \zeta^{2} \, a_2$, and employ equation (\ref{adiabatapprox}) to obtain an expression for the adiabatic threshold: 
\begin{align}
\label{adiabatTypeI}
&\frac{M_{\rm{disk}}}{\M} \lesssim \frac{4}{s} \left( \frac{h}{a} \right)^2 (k-1)^{11/9} \nonumber \\  
&\times \frac{3^{2/3} \left( f_{\rm{res}}^{(1)} (m_1 + m_2)  \right)^{4/3} }{\left( \left( (k-1)^2 \, k\right)^{1/3} m_1 - (k -1 ) m_2 \right) \M^{1/3}},
\end{align}
where $M_{\rm{disk}} = 2 \pi \, \Sigma_0 \, a_0 \, a_2$ is the disk mass contained interior to the outer planetary orbit\footnote{In this definition, it is implicitly assumed that $a_2$ greatly exceeds the truncation radius of the disk.}. This relationship is depicted for a particular mass ratio and disk aspect ratio in Figure (\ref{typeIfig}).

It is worth noting that while the resonant dynamics themselves tend to only depend on the cumulative planetary mass, the above expression is indeed strongly dependent on the mass ratio. The additional dependence on the disk surface density profile and separation from the central star yields a diverse range of physical parameters which may fall either into the strongly adiabatic or strongly non-adiabatic regimes (see e.g. \citealt{PapaloizouSzuszkiewicz2005,CresswellNelson2008,Ketchum2011}).

Having considered convergent migration of sub-Jovian planets, let us now examine the case where the planetary mass is sufficient to gravitationally clear out a substantial gap around its co-orbital neighborhood. In this situation, the planet is shepherded towards the central part of the gap where all torques instantaneously cancel. Maintaining this configuration, the planet drifts inwards along with the accretionary flow of the gas \citep{LinPaploizou1986,CridaMorby2007,MorbyCrida2007}. Therefore, the orbital evolution rate in this so-called "type-II" regime is largely controlled by the global angular momentum transport within the disk\footnote{Strictly speaking, this is only true if the planet mass is not large enough to disrupt the process of viscous accretion. In other words, the dominant portion of the angular momentum budget within the planet's neighborhood must reside in the disk material. Quantitatively, this criterion is satisfied when $m \lesssim 4 \pi \, \Sigma \, a^2$, where $\Sigma$ is the gas surface density immediately outside the gap \citep{Baruteau2014}.}, and correspondingly operates on the viscous timescale:
\begin{align}
\label{typeII}
\tau_{\rm{II}} = - \frac{2}{3} \frac{a^2}{\nu} = - \frac{2}{3}\frac{1}{\bar{\alpha}} \left(\frac{h}{a} \right)^{-2} \frac{1}{n}.
\end{align}
Here, $\nu$ is the kinematic viscosity of the gas \citep{Armitage2010}. 

\begin{figure}
\includegraphics[width=0.97\columnwidth]{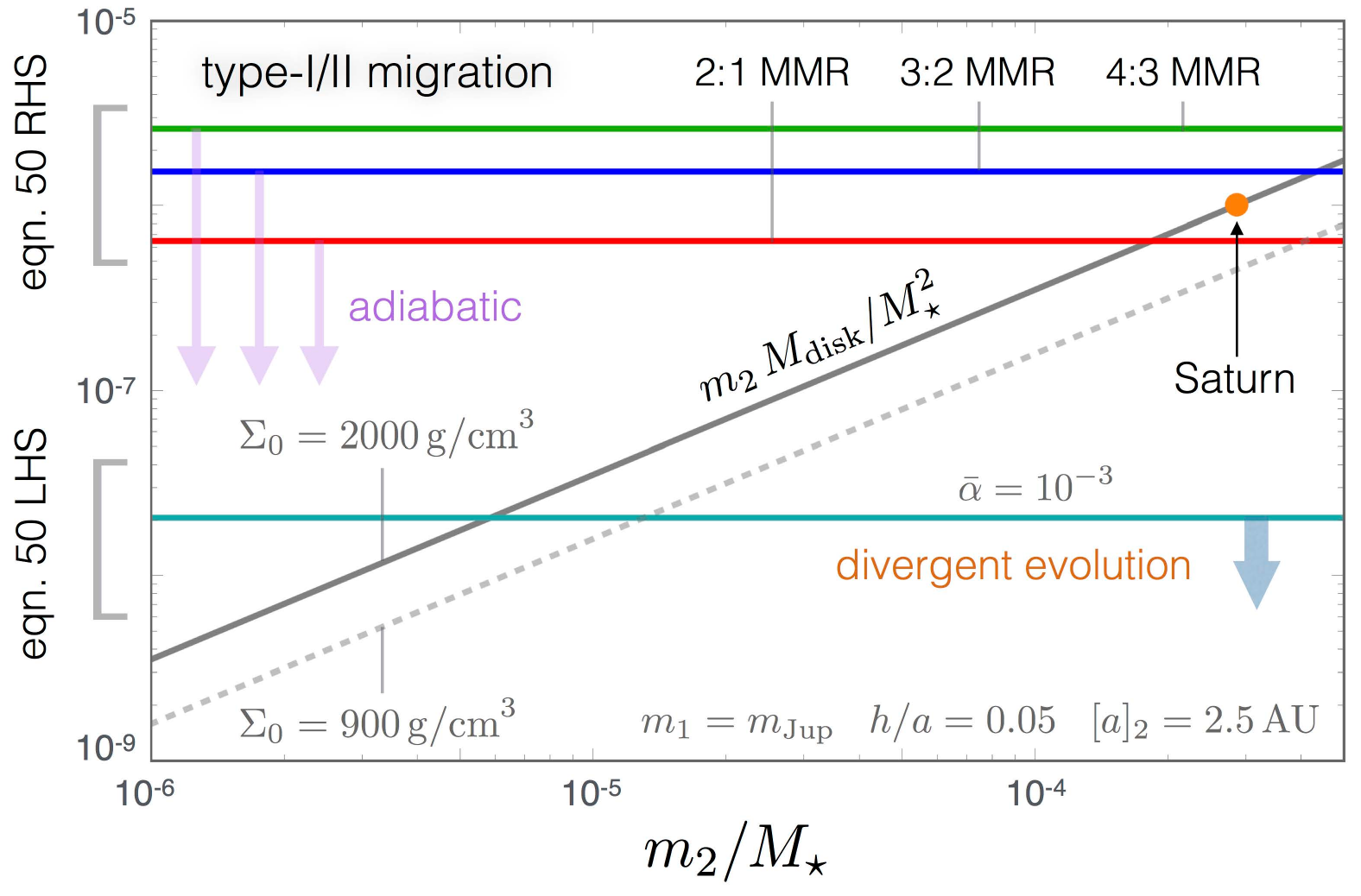}
\caption{The adiabatic threshold for two objects, where the inner planet has the planet-star mass ratio identical to Jupiter and migrates in the type-II regime, while the outer planet of mass $m_2$ undergoes type-I migration in an isothermal Mestel-like disk. The red, blue, and green curves labeled by resonant period ratios correspond to the adiabatic criteria and are given by the RHS of equation (\ref{typeIIconv}). The intersection between the cyan line (given by the LHS of equation \ref{typeIIconv}) and the gray curve of slope 2 denotes the minimum mass, below which type-II migration proceeds at a faster rate than type-I migration, leading to divergent evolution. The assumed disk parameters are quoted in the figure and roughly mimic those of the minimum mass solar nebula. Note that for the utilized fiducial parameter choices, Saturn fails the adiabatic criterion for the 2:1 resonance with Jupiter, but satisfies the threshold that corresponds to the 3:2 resonance. However, capture into the 2:1 resonance is possible, given a diminished surface density profile.}
\label{typeI_IIfig}
\end{figure}

Equation (\ref{typeII}) implies that within the context of the envisioned (simplified) picture, two planets that are simultaneously undergoing type-II migration will never experience convergent resonant encounters. Instead, convergent evolution is only possible if a less massive object (that migrates in the type-I regime) on an external orbit catches up to a more massive, slowly migrating planet. For this scenario to hold true, the outer planet ($m_2$) must fail the gap-opening criterion (\ref{gapopen}).

For resonant capture to be successful within the framework of this setup, the system must satisfy two additional criteria. Specifically, the outer body must not be so small that its type-I orbital decay proceeds at a slower rate than type-II migration of the inner body. Simultaneously, the outer body must not be massive enough for its migration rate to overwhelm the adiabatic limit (\ref{adiabaticcrit}). These two requirements bracket the product of $m_2$ and $M_{\rm{disk}}$ from below and above:
\begin{align}
\label{typeIIconv}
&\frac{3 \pi}{s}\frac{k}{k-1} \left( \frac{h}{a} \right)^4 \bar{\alpha} \lesssim \bigg( \frac{m_2}{\M} \bigg) \bigg( \frac{M_{\rm{disk}}}{\M} \bigg) \nonumber \\
&\lesssim 4 \frac{ \left((27(k-1)\right)^{2/9}}{s} \bigg(\frac{f_{\rm{res}}^{(1)} \, m_1}{\M} \bigg)^{4/3} \bigg( \frac{h}{a} \bigg)^2.
\end{align}
A graphic representation of this criterion for a Jupiter-mass inner planet is shown in Figure (\ref{typeI_IIfig}).

In obtaining equation (\ref{typeIIconv}), we have made a series of approximations. In particular, we used a simplified form of the adiabatic threshold as before. Subsequently, we assumed that the majority of the planetary mass is contained within the inner body. Finally, for the sake of a less cumbersome expression, we assumed that convergent evolution is dominated by type-I migration of the outer body in the second inequality. These simplifications generally hold as long as the bracketed quantity is not too close to either limit.

For typical disk parameters (e.g. $h/a \sim 0.05$; $\bar{\alpha} \sim 0.001$), the extrema in equation (\ref{typeIIconv}) differ by a little more than an order of magnitude. However, it should be noted that without invoking unreasonable quantities, it is possible to make the range of planetary masses for which adiabatic capture may take place, exceptionally small. 

\smallskip
$\bullet$ \textit{Planetesimal-Driven Migration}
\smallskip

In the post-nebular stage of planetary system evolution, torques derived from interactions with the gas are non-existent. However, large-scale migration can still occur as a consequence of asymmetric scattering of planetesimals by a more massive object \citep{FernandezIp1984,Malhotra1995}. Within the context of the early dynamical evolution of the outer Solar System, this process may have played a crucial role, as scattering-facilitated divergent evolution of Jupiter and Saturn has been invoked as a means of igniting a transient dynamical instability\footnote{Among the many features whose origins can be attributed to this instability are the so-called period of late heavy bombardment \citep{Gomes2005,Levison2011}, Jupiter's Trojan asteroids \citep{Morbidelli2005}, irregular satellite populations of the giant planets \citep{Nesvorny2007}, and the dynamical structure of the Kuiper belt \citep{Levison2008,BatyginBrown2011}.} \citep{Tsiganis2005,BatyginBrown2010}. 

\begin{figure}
\includegraphics[width=1\columnwidth]{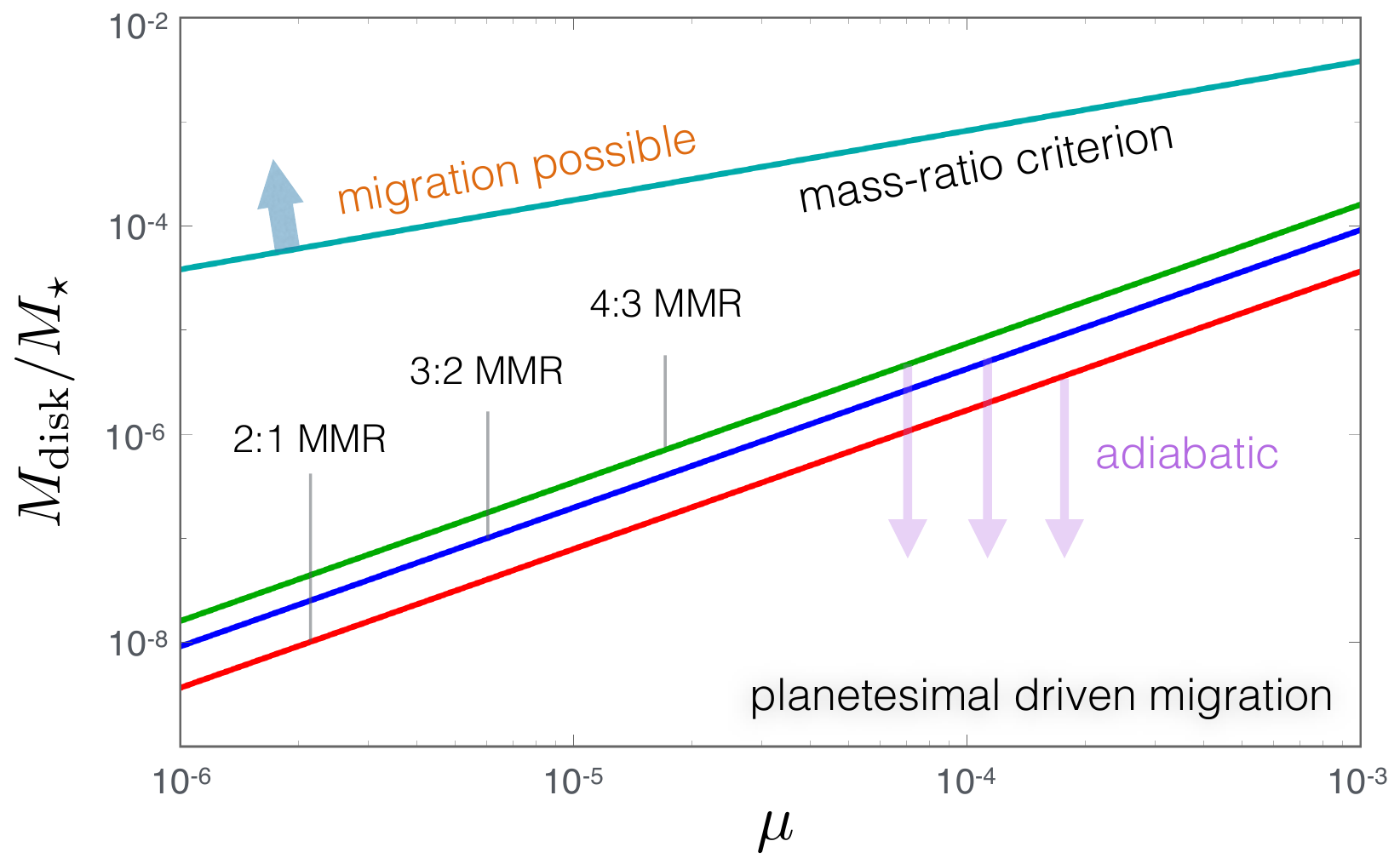}
\caption{The adiabatic threshold for planetesimal driven migration. As in previous figures, the adiabatic criteria are shown with red, blue, and green lines. However, these curves lie well below the mass-ratio criterion (shown with a cyan line) which must be exceeded for migration to be self-sustaining. This means that planetary embryos that migrate through long-lived planetesimal disks generally do not suffer adiabatic resonant encounters. }
\label{pldrfig}
\end{figure}

A prominent setting where planetesimal-driven migration can lead to substantial orbital changes is a gas-depleted planetesimal disk with embedded planetary embryos. In light of this, it is worthwhile to examine if resonant locking among such embryos can occur. A fiducial timescale on which planetesimal-driven migration operates is \citep{Ida2000,Kirsh2009}
\begin{align}
\label{tauPDM}
\big| \tau_{\rm{P}} \big| = \frac{\M}{2 \, \Sigma \, a^2} \frac{1}{n},
\end{align}
where $\Sigma$ now refers to planetesimal surface density, for which we have once again assumed a Mestel-like radial profile. A detailed study of planetesimal-driven migration was recently carried out by \citet{MintonLevison2014}, who identified five criteria that must be satisfied for this mode of orbital transport to be self-sustained. Here, we shall not dwell extensively on these requirements, but simply assume that they are satisfied for at least one of the secondary objects of interest.

The direction of planetesimal-driven migration can be inward or outward. To this end, \citet{Kirsh2009} find a preference for orbital decay over growth, however \citet{MintonLevison2014} find no statistically significant predisposition for either direction. From equation (\ref{tauPDM}), it is clear that the inner orbit changes on a shorter timescale. Therefore, if both orbits decay at rates given by equation (\ref{tauPDM}), the evolution is bound to be divergent. Reversing this argument, one may envisage that convergent evolution is possible if both orbits grow. However, in this case the inner planet would be migrating in the wake of the outer planet, through a pre-excited disk. Without invoking very rapid dissipation that would dynamically cool the disk, such a scenario would likely violate the disk eccentricity criterion (see \citealt{MintonLevison2014}), halting the inner orbit's migration. Consequently, it would appear that given nominal parameters, convergent evolution is possible either if the inner orbit evolves outwards while the outer evolves inwards, or if one of the orbits remains stationary while being approached by the other.

In principle, each of these situations will be characterized by its own, subtly different adiabatic threshold. However, in interest of succinctness we shall only consider the case where both orbits actively approach one-another, noting that the corresponding expressions for the case where one of the planets remains stationary are quantitatively similar, to within a factor of $\sim 2$. Combining expressions (\ref{tauPDM}) and (\ref{adiabaticcrit}), we have: 
\begin{align}
\label{PDMadcrit}
\frac{M_{\rm{disk}}}{\M} &\lesssim \frac{3^{2/3}}{\pi} \frac{(k-1)^{11/9}}{k-1+(k\,(k-1)^2)^{1/3}} \nonumber \\
&\times \bigg( \frac{ f_{\rm{res}}^{(1)} (m_1 + m_2)}{\M}  \bigg)^{4/3}.
\end{align}

It is interesting to note that while expression (\ref{PDMadcrit}) provides an upper bound on the disk mass, the mass ratio criterion for operation of planetesimal-driven migration requires the mass of the migrating body to not exceed its encounter mass (equivalently, isolation mass; \citealt{Lissauer1987}) by more than a factor of $\sim 3$ \citep{Kirsh2009}, thereby providing a lower bound. Given that the multiplicative factor of $(m_1+m_2)/\M$ is of order unity for all resonances of interest and that the stopping mass is approximately $(m/\M)_{\rm{stop}} \simeq 12 \, (M_{\rm{disk}}/\M)^{3/2}$, we can formulate the following rough criterion for scattering-facilitated adiabatic resonant encounters:
\begin{align}
\label{PDMadcritapprox}
\frac{M_{\rm{disk}}}{\M} \lesssim \bigg(\frac{m_1 + m_2}{\M}  \bigg)^{4/3} \lesssim 27 \, \bigg( \frac{M_{\rm{disk}}}{\M} \bigg)^2.
\end{align}

Despite its crudeness, the above expression is informative, and is shown in Figure (\ref{pldrfig}). In particular, it dictates that in order to simultaneously satisfy the adiabatic criterion and the mass ratio criterion, the planetesimal disk mass must exceed $M_{\rm{disk}}/\M \gtrsim 0.04$. Such a mass ratio is characteristic of gaseous protoplanetary disks, and thus exceeds the mass ratio inherent to the solid component of the disk by approximately two orders of magnitude. Therefore, equation (\ref{PDMadcritapprox}) implies that without appealing to special configurations (such as the primordial multi-resonant state of the outer Solar System), planetary migration facilitated by scattering of planetesimal cannot lead to convergent adiabatic resonant encounters.

\smallskip
$\bullet$ \textit{Tidal Evolution}
\smallskip

A final mode of convergent migration that we shall consider here is that facilitated by tidal dissipation. This mechanism is particularly relevant to planetary satellites, as orbital changes induced by tides raised on the host planet are considered to be the dominant driving mechanism\footnote{See however, \citet{PealeLee2002} and \citet{CanupWard2002} for an alternative, circumplanetary disk driven view on the assembly of the Laplace resonance among the Galilean satellites.} responsible for resonant pairs of satellites in the Solar System \citep{Goldreich1965,Peale1976,Peale1986,Henrard1983}. Correspondingly, we shall frame the following discussion in the satellite-planet context, keeping in mind that similar arguments can apply to close-in planets and their host stars (see e.g. \citealt{AdamsBloch2015}).

Following the works of \citet{Allan1969} and \citet{Dermott1988}, we shall assume that changes in the semi-major axis induced by satellite tides can be neglected in favor of their planetary counterparts. In this case, the characteristic migration timescale for an orbit residing beyond the synchronous radius is given by \citep{MD99}:
\begin{align}
\label{tautide}
\tau_{\rm{T}} = \frac{1}{3} \frac{Q}{k_2} \frac{M_{\rm{p}}}{m} \left(\frac{a}{R_{\rm{p}}} \right)^5  \frac{1}{n},
\end{align}
where $Q$ is the specific dissipation function, $k_2$ is the Love number, and $R_p$ is the physical radius of the planet.

\begin{figure}
\includegraphics[width=1\columnwidth]{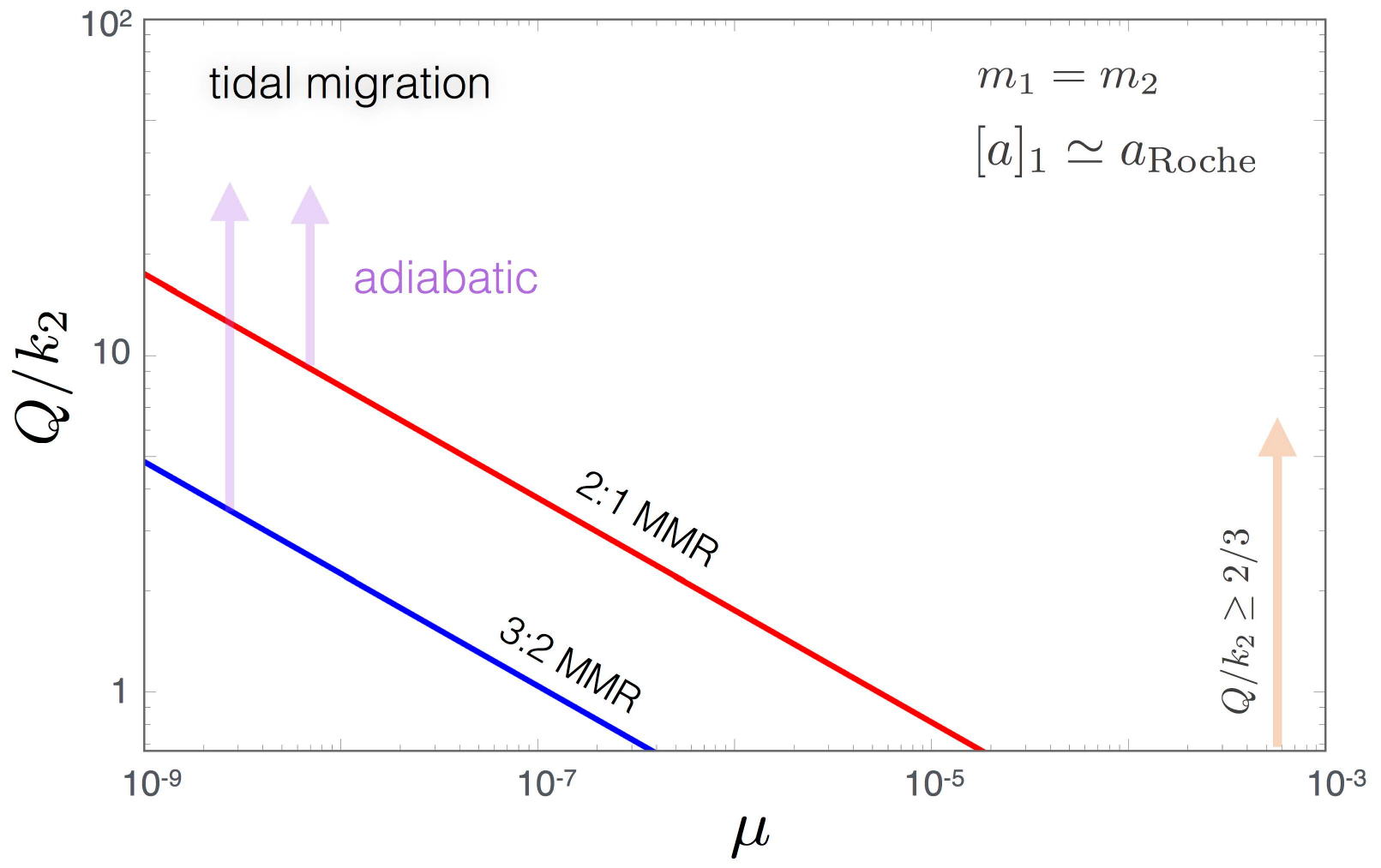}
\caption{The adiabatic threshold corresponding to tidal evolution of an equal-mass pair of satellites. For illustrative purposes, in this figure we considered the inner object's orbit to approximately coincide with the Roche radius, yielding a maximal migration rate. Even in this extreme case, an reasonable dissipation efficiency will yield adiabatic evolution.}
\label{tidalfig}
\end{figure}

Contrary to the case of type-I migration discussed above, the expression (\ref{tautide}) implies that the outer satellite mass cannot exceed the inner satellite mass by more than $m_2 < m_1(k/k-1)^{13/2}$ for evolution to remain convergent. Substitution of the tidal timescale into equation (\ref{adiabatapprox}) yields the following expression for the adiabatic threshold:
\begin{align}
\label{tideadiabat}
\frac{Q}{k_2} \left( \frac{a_2}{R_{\rm{p}}} \right)^5 &\gtrsim \frac{3^{1/3} \pi}{2} \left( \frac{M_{\rm{p}}}{ f_{\rm{res}}^{(1)} (m_1 + m_2) }  \right)^{4/3} \nonumber \\
&\times \left( \frac{k^{13/2} m_1 - (k-1)^{13/2} m_2}{(k-1)^{41/9} M_{\rm{p}}} \right).
\end{align}

To evaluate the physical meaning of equation (\ref{tideadiabat}), it is instructive to consider a limiting parameter regime that maximizes the migration rate. Accordingly, let us envision that the total secondary mass is concentrated entirely in the inner satellite which orbits immediately outside the Roche limit (i.e. $a_2 \simeq (5/2)\, \zeta^{2} \, R_{\rm{p}}$), and the planet is a homogenous fluid body with a Love number of $k_2 = 3/2$. Setting the specific dissipation function to a minimal\footnote{The smallest possible value that $Q$ can take on is unity. Physically, this would correspond to complete dissipation of all energy stored within a single tidal cycle. However, for the weak-friction theory (within the context of which equation (\ref{tautide}) is obtained) to apply, $Q$ must greatly exceed unity \citep{Hut1981}.} value of $Q \sim 10$, we can obtain a critical mass ratio $(m_1/M_{\rm{p}})_{\rm{crit}}$ below which the adiabatic limit is broken. Upon evaluation, we obtain $(m_1/M_{\rm{p}} )_{\rm{crit}}\sim 10^{-7},10^{-8}$, and $10^{-9}$ for the 2:1, 3:2, and 4:3 resonances respectively.

While these values already lie well below the typical range of natural\footnote{Note that the assumptions of low eccentricity and inclination inherent to the formulation of the Hamiltonian (\ref{Horbel}) render our treatment largely inapplicable to irregular satellites of the Solar System.} satellite - planet mass ratios, it is important to keep in mind that we have invoked unreasonably rapid tidal evolution (as an example, for Jupiter, $k_2/Q \sim 10^{-5}$ whereas we have used $k_2/Q \sim 3/20$; \citealt{GoldreichSoter1966,Lainey2009}) in order to derive them. In other words, the obtained quantities are gross overestimates of the practically relevant critical mass ratios. Recall also that the rate of tidal evolution diminishes with $(m/M_{\rm{p}})$ meaning that the critical mass ratio may also lie in a regime where $\tau_{\rm{T}}$ greatly exceeds the age of the system, rendering this quantity meaningless. Therefore, for all practical purposes it is likely safe to assume that tidally-facilitated resonant encounters will almost always lie in the adiabatic regime. For completeness, Figure (\ref{tidalfig}) shows the adiabatic threshold for an equal-mass pair of satellites. 

\smallskip

The calculations presented above yield a series of practically useful criteria that inform whether or not convergent migration within a given physical setting adheres to the adiabatic limit. In interpreting this discussion however, it is important to keep in mind that for the sake of definitiveness, we have limited our calculations to simplified formulae that describe nominal dissipative evolution rates of planets. In reality, the detailed physics of disk-planet interactions, planetesimal scattering and tidal evolution can be rather complex (see \citealt{KleyNelson2012,Baruteau2014,MintonLevison2014,EfroimskyWilliams2009} for in-depth discussions). Indeed, substantial deviations from the quoted prescriptions can occur within a more comprehensive treatment. As a result, the derived criteria should be viewed as approximate, rather than conclusive.

\subsection{Specific Applications}

\subsubsection{Resonant Capture of Jupiter and Saturn in the Protosolar Nebula}

As already mentioned above, type-II migration generally causes planets to spiral in towards their host stars. In light of this, the following question naturally arises: why are Jupiter and Saturn where they are today? As an answer to this question, it was shown by \citet{MassetSnellgrove2001} that the simultaneous dynamics of Jupiter and Saturn, when submerged in the protosolar nebula, alter the isolated migration picture qualitatively. In particular, simulations show that Jupiter and Saturn exhibit convergent orbital evolution, and eventually get captured in resonance. With a resonant lock established, the planets carve out a mutual gap, which (owing to a particular planet-planet mass ratio) alters the torque balance of the system and facilitates joint outward migration of the resonant pair \citep{MorbyCrida2007}. 

Although resonant reversal of orbital decay was initially proposed to account for the retention of Jupiter and Saturn at large orbital radii, such a sequence of events smoothly connects the nebular stage of Solar System evolution to the compact initial conditions of the Nice model \citep{Morbyetal2007,BatyginBrown2010,NesvornyMorbidelli2012}, and provides natural explanations for the comparatively small mass of Mars \citep{Walsh2011}, the composition of the Asteroid belt \citep{OBrien2014}, and the non-existence of otherwise common close-in Super-Earths inside of Mercury's orbit \citep{BatyginLaughlin2015}. Moreover, it has been proposed that a similar mechanism is ubiquitously responsible for halting large-scale migration of giant exoplanets that reside at wide orbital separations \citep{Morbidelli2013}. 

Following the pioneering study of \citet{MassetSnellgrove2001}, early dynamical evolution of Jupiter and Saturn has been studied by \citet{MorbyCrida2007,PierensNelson2008,DAngeloMarzari2012,Pierens2014} using hydrodynamical simulations. In agreement with the initial experiments of \citet{MassetSnellgrove2001}, most studies find that independent of initial conditions, the common outcome is passage through the 2:1 resonance and capture into the 3:2 resonance. The prevalent speculation (and indeed this turns out to be correct) within the aforementioned works is that failure of the 2:1 resonance to capture Jupiter and Saturn arises from non-adiabatic evolution. Given the well-defined nature of the problem, it is instructive to examine how the results of numerical simulations fit into the framework of the developed analytic theory.

As a first step, let us imagine that the adiabatic criterion is satisfied. In this regime, we can use equation (\ref{criterion}) to inform the conditions under which resonant capture is not guaranteed. Specific curves corresponding to the 2:1 Jupiter-Saturn resonance are shown in Figure (\ref{JS_21fig}). Evidently even in the conservative case, pre-encounter eccentricities in excess of $e \gtrsim 0.1$ are required to reduce the adiabatic capture probability below $\sim75\%$. Conventional theory (e.g. \citealt{GoldreichTremaine1980,Ward1997}) suggests that dissipative evolution of planets within axisymmetric disks tends to damp orbital eccentricities and inclinations\footnote{See \citet{GoldreichSari2003,Tsang2014} for an alternative view.}, meaning that capture into the 2:1 resonance is highly likely if the adiabatic limit is satisfied. Recall also that the 2:1 resonance allows for guaranteed capture at higher eccentricity than the 3:2 resonance, meaning that if the system fails to capture into the 2:1 resonance due to being too dynamically pre-excited, it will also likely fail to capture into the 3:2 resonance. 

Let us now evaluate the critical rates of convergent evolution for Jupiter and Saturn in the protosolar nebula, as dictated by equation (\ref{typeIIconv}). Adopting the typically quoted aspect ratio of $h/a = 0.05$, a viscosity parameter of $\bar{\alpha} = 10^{-3}$, a surface density of $\Sigma_0 = 2000$ g/cm$^2$ at $a_0 = 1$ AU, and an encounter semi-major axis of $a_2 \simeq 2.5$ AU (the envisioned ``tacking" radius), the minimum mass of the outer planet above which the orbits will approach each-other is $m_2 \gtrsim 3 M_{\oplus}$. This is in excellent agreement with the simulation results of \citet{PierensNelson2008} who find the minimum mass to be $m_2 \gtrsim 3.5 M_{\oplus}$. 

Using the same expression we find the threshold mass, above which the adiabatic criterion is violated for the 2:1 resonance, is $m_2 \gtrsim 60 M_{\oplus}$, or about $2/3$ of Saturn's mass. For the 3:2 resonance, the critical value increases to $m_2 \gtrsim 140 M_{\oplus}$, which comfortably exceeds Saturn's mass. Again, these estimates are in perfect alignment with the numerical experiments of \citet{PierensNelson2008} who observe capture into the 2:1 resonance for $m_2 = 30 - 40 M_{\oplus}$ and capture into the 3:2 resonance for $m_2 = 80 - 100 M_{\oplus}$. Thus, our analytical theory robustly conforms to the results of numerical experiments and analytically demonstrates that the recurrent capture of Saturn into a 3:2 rather than a 2:1 mean motion resonance is indeed an outcome of the violation of the adiabatic threshold for the 2:1, but not the 3:2 resonance.

Equation (\ref{typeIIconv}) further implies that capture into the 2:1 resonance is indeed possible, provided a reduced disk surface density. As an example, in addition to results corresponding to fiducial parameters quoted above, Figure (\ref{typeI_IIfig}) also shows a supplementary (dashed) curve that represents a disk with $\Sigma_0 = 900$ g/cm$^2$. The fact that capture is permitted in this regime is fully consistent with the simulation results of \citet{Pierens2014}, who numerically obtain capture into the 2:1 resonance for sub-nominal surface densities. 

\begin{figure*}
\includegraphics[width=1\textwidth]{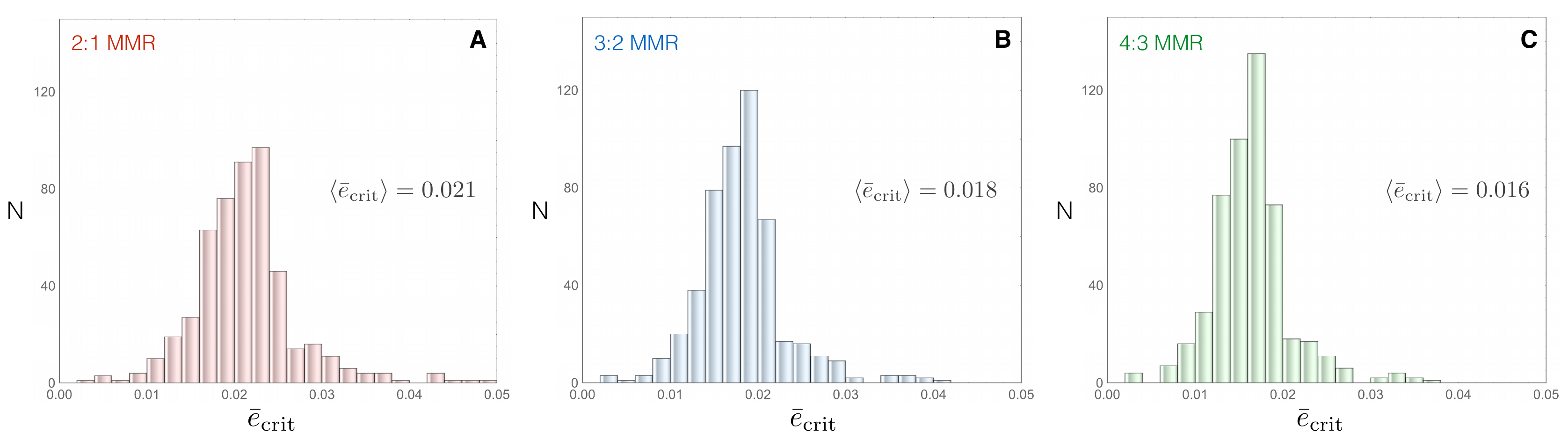}
\caption{Distributions of critical orbital eccentricities, above which resonant capture of confirmed planets pairs of the \textit{Kepler} survey is not guaranteed. Because capture probability decreases rapidly with pre-encounter orbital excitation (see Figure \ref{pcapt}B), the quoted average eccentricities are approximately indicative of the true eccentricities at which the outcomes of adiabatic resonant encounters switch from preferential capture to preferential transition across the resonance. Results presented in panels A, B, C correspond to 2:1, 3:2, 4:3 mean motion resonances respectively.}
\label{ecritfig}
\end{figure*}

Naturally, the analytic model developed herein does not take into account the full richness of the possibilities that can occur in real protoplanetary disks. For example, simulations show that after Saturn passes through the 2:1 resonance, it directly invades the disk gap opened by Jupiter. As a result, the diminished local surface density of the gas further reduces its migration rate, allowing for an even more robust satisfaction of the adiabatic criterion \citep{MorbyCrida2007}. A similar reduction in Saturn's migration rate can in principle occur if Saturn successfully opens its own gap, such that its migration rate is still faster than the strict type-II rate given by equation (\ref{typeII}), but only slightly so. As is made clear by equation (\ref{gapopen}), this can be achieved by reducing the aspect ratio or the viscosity of the disk.

\subsubsection{Paucity of Resonances Among Close-In Sub-Jovian Planets}

Having examined the well-studied problem of nebular evolution of Jupiter and Saturn, let us now consider a somewhat less transparent phenomenon. A pivotal realization that has emerged as a consequence of the recent identification of thousands of extrasolar planets by the \textit{Kepler} transit survey, is that the dominant mode of planet formation within the Galaxy is one that produces planets substantially smaller than Jupiter and Saturn with orbital periods below $\sim 100$ days \citep{Youdin2011,Howard2012}. These close-in sub-Jovian planets are estimated to orbit approximately half of all Sun-like stars, and often comprise tightly packed multi-planet systems \citep{Mayor2011,Batalha2013,Lissauer2014}. Accordingly, understanding the architectural origins of such systems holds the key to identifying the dominant processes at play during the epoch of planet formation.

The orbital distribution of close-in sub-Jovian planets is for the most part, devoid of orbital resonances. While there exist substantial enhancements in the periods ratios immediately outside of the 2:1 and 3:2 commensurabilities (and such systems can indeed be attributed to resonant evolution in presence of dissipative forces; \citealt{LithwickWu2012,Delisle2012,BatyginMorby2013}), these planet-pairs comprise a minority within the observational sample \citep{Fabrycky2014}. This fact is puzzling when viewed within the context of the theoretical expectation that type-I migration should typically generate crowded resonant chains in protoplanetary disks \citep{CresswellNelson2008}.

One way to prevent resonant capture is to invoke a sufficiently massive disk such that the adiabatic criterion inherent to type-I migration (equation \ref{adiabatTypeI}) is not satisfied \citep{Quillen2006,MustillWyatt2011}. Although this will indeed provide an effective avenue towards halting the production of resonant chains of planets, this process alone is unlikely to explain the observed exoplanet architectures in a satisfactory manner for two reasons. First, the disk mass itself decreases in time, generally on a timescale that is much longer than typical type-I migration timescales \citep{Haisch2001,HerczegHillenbrand08}. This means that even if a given disk starts out in a state where resonant capture is hampered by overly rapid migration, such a disk will eventually evolve to a state where the resonant condition (\ref{adiabatTypeI}) is satisfied. Second, it has been shown that even in massive disks, type-I migration is not ubiquitously fast, as suggested by conventional linear calculations \citep{Paardekooper2010,Paardekooper2011,Bitsch2011,Bitsch2013}. The main deviation from linear theory arises from the fact that entropy gradients in radiative disks alter the disk-planet interaction in such a way as to create zones where type-I migration greatly slows down, reverses, or ceases all together, allowing for convergent congregation of low-mass planets \citep{McNeil2005,PapaloizouSzuszkiewicz2005,Matsumoto2012}.

An alternative means of generating a largely non-resonant orbital distribution is to invoke processes that tend to destabilize mean-motion commensurabilities. To this end, \citet{Adams2008}, followed by \citet{Ketchum2011} and \citet{Rein2009} proposed that stochastic perturbations due to turbulent forcing within protoplanetary disks may inhibit long-lived resonances (see also \citealt{Laughlin2004} for a related discussion). Another possibility, recently put forward by \citet{GoldreichSchlichting2014} is that the combined effects of semi-major axis and eccentricity damping may render resonant configurations unstable on long timescales. While both of these proposals are in principle reasonable, they ultimately rely on processes that are poorly constrained. Specifically, these mechanisms inherently depend on quantities such as the duty-cycle of turbulent eddies and the relative damping rates of orbital eccentricities and semi-major axes, both of which remain scantily understood and constitute active fields of research \citep{BaiStone2013,Bitsch2015}. As a result, in the spirit of Occam's razor, it is tempting to inquire if a simpler mechanism that can preclude resonant capture exists.

Keeping in mind the arguments presented above, let us only consider the adiabatic regime. As we already saw in the beginning of this section, capture is guaranteed only below some critical extent of pre-encounter orbital excitation, and capture probability decreases rapidly above this value (Figure \ref{pcapt}B). How eccentric must have close-in planet pair been to preferentially skip over resonances without locking? To obtain a rough answer to this question, consider the following calculation. 

Suppose that by default, planets in protoplanetary disks migrate not on exactly circular orbits but with some small eccentricities, $\bar{e}$. Further, let us further speculate that $\bar{e}$ will be similar for two planets that are about to encounter a mean motion resonance, as it is set by external factors rather than direct planet-planet interactions. Provided that planetary and stellar masses are known, the critical value of pre-encounter eccentricity, $\bar{e}_{\rm{crit}}$, can be computed by means of equation (\ref{criterion}), setting $\Delta \varpi = \pi$. We have performed this calculation for all confirmed planet-pairs detected by the $\textit{Kepler}$ transit survey.

Figure (\ref{ecritfig}) shows the distributions of $\bar{e}_{\rm{crit}}$ corresponding to the 2:1, 3:2 and 4:3 resonances for $\textit{Kepler}$ planets. Clearly, the pre-encounter orbital eccentricities required to render capture uncertain are ubiquitously low: $\bar{e}_{\rm{crit}} \simeq 0.01-0.03$. More importantly however, our simple estimate for $\bar{e}_{\rm{crit}}$ is astonishingly close to the observed rms eccentricities of the $\textit{Kepler}$ sample. Specifically, from their analysis of transit timing variations, \citet{WuLithwick2013} and \citet{HaddenLithwick2014} find that three quarters of the considered sub-sample of planets conforms to an eccentricity distribution with an rms free eccentricity of $e \sim 0.02$, while the remaining quarter is characterized by substantially higher values. Although caution must be exercised in interpreting the present eccentricities of sub-Jovian planets as having been inherited from their natal disks, the conspicuous similarity of the theoretically required values to the corresponding observations is suggestive, and opens a previously unexplored avenue for explaining the origins of the orbital distribution.

\section{Discussion}

Both theoretical and observational lines of inquiry suggest that mean-motion resonances play a central role in the formation and long-term evolution of planetary systems (see e.g. \citealt{Morbyetal2007,Rivera2010,Deck2012,GoldreichSchlichting2014} and the reference therein). While this fact has been well recognized as a consequence of countless numerical experiments \citep{Quillen2006,Ketchum2011,OgiharaKobayashi2013}, until now a comprehensive qualitative understanding of the conditions for capture of planets into resonances had remained elusive. In this work, we have taken steps towards formulating a model for capture into resonance in a generic, analytical way. Our development provides an underlying framework within which the outcomes of numerical experiments can be interpreted. Moreover, we provide a series of simple criteria which can be used to inform the outcome of dynamical simulations. 

As a practical recipe for theoretical analysis of resonant encounters, we propose the following order of calculation: \\
$\bullet$ Does the convergent evolution rate of the system satisfy the adiabatic condition? For the specific cases of orbital migration driven by interactions with a gaseous disk, scattering of planetesimals, or tidal evolution, expressions (\ref{adiabatTypeI}) or (\ref{typeIIconv}), (\ref{PDMadcritapprox}), and (\ref{tideadiabat}) may be used respectively. Alternatively, the generalized adiabatic criterion (\ref{adiabaticcrit}), or its simplified form (\ref{adiabatapprox}) can be employed. If the criterion is violated, capture will not take place. \\
$\bullet$ If the resonant encounter lies in the adiabatic regime, is capture certain? This can be assessed using expression (\ref{criterion}). If knowledge of the differences in the apsidal lines, $\Delta \varpi$, is available, the guaranteed capture equation can be used directly. If $\Delta \varpi$ is not known, it is sensible to assume $\Delta \varpi = \pi$, as this yields the most conservative estimate.\\
$\bullet$ If adiabatic capture is not guaranteed, what is the capture probability? If $\Delta \varpi$ is known, the initial condition far away from resonance can be related to the location of the unstable fixed point of the resonance at separatrix-crossing using expression (\ref{areamatch}). The corresponding probability can then be calculated using equation (\ref{Pcapt}). If $\Delta \varpi$ is unspecified, the averaged capture probability given by equation (\ref{Pcaptave}) can be calculated.

In this work, we have utilized the developed formalism to consider two specific applications. First, we analyzed the numerically well modeled evolution of Jupiter and Saturn's orbits in the primordial Solar nebula (\citealt{Pierens2014} and the references therein). Within the context of this problem, our theoretical arguments confirmed the previously insinuated notion (e.g. \citealt{MorbyCrida2007}) that the tendency of Jupiter and Saturn to lock into the 3:2 rather than 2:1 resonance is a consequence of Saturn's rapid migration and the associated violation of the adiabatic criterion. We subsequently considered the origins of the dominantly non-resonant orbital distribution of close-in sub-Jovian multi-planet systems discovered by $\textit{Kepler}$ \citep{Batalha2013,Fabrycky2014}. Specifically, we showed that if low-mass planets generally reside on slightly eccentric orbits (i.e. $e\gtrsim0.02$) when submerged in their natal disks, the chances of resonant capture are greatly diminished.

Although the theoretically derived threshold eccentricities (below which resonant capture is hindered) are almost identical to the present observationally inferred values within the $\textit{Kepler}$ sample \citep{WuLithwick2013}, we have not specified the physical origin of these subtle deviations from circular orbits. In principle, there exists a large number of dynamical mechanisms that may lead to such excitation. Among them are orbital excitations by turbulent forcing \citep{NelsonPapaloizou2004,AdamsBloch2009} and interactions with distant massive planets which may themselves experience violent dynamical instabilities within protoplanetary nebulae \citep{Lega2013}. 

An arguably more intriguing idea is that steady-state non-zero eccentricities may stem directly from the interactions of planets with their natal disks\footnote{Note that the timescale of eccentricity modulation due to planet-disk interaction is almost certainly much shorter than that corresponding to semi-major axis evolution \citep{GoldreichTremaine1980,LeePeale2002}.}. A body of recent literature has shown that contrary to the conventional theoretical simplification of perfectly circular protoplanetary nebulae, real disks exhibit substantial deviations away from axial symmetry and are indeed believed to be globally lopsided \citep{2009ApJ...704..496B,2012A&A...547A..84T,2013Natur.493..191C,2013PASJ...65L..14F,2013ApJ...775...30I,2013Sci...340.1199V,2014ApJ...783L..13P,2014A&A...562A..26B}. Although observations taken in the dust continuum wavelength band severely overrepresent the corresponding non-axisymmetric over-densities of gas (which holds the vast majority of the disk mass), the gas disk eccentricities\footnote{Possible mechanisms that may be responsible for maintenance of global lopsided modes of disks include disk self-gravity (\citealt{2008MNRAS.387....2D,2015ApJ...798L..25M}; see also \citealt{Touma2009} and \citealt{Batygin2012} for a related discussion) and hydrodynamic forces \citep{Larwood1996,2014MNRAS.440.1179X}. Moreover, excitation of disk eccentricities may arise from external perturbations by passing or bound stars in star formation environments that are well-known to exhibit enhanced stellar multiplicity \citep{2013ARA&A..51..269D}.} required to explain the ALMA observations \citep{2015ApJ...798L..25M} may very well be sufficient to also perturb low-mass planets onto slightly non-circular orbits. While in this work, we have taken initial steps towards exploring this idea, undoubtably much additional effort is required to quantitatively assess the viability of this hypothesis. 

Our analytical study of resonant capture complements a number of recent developments that employ the same integrable formalism for resonant dynamics. In a closely related study \citep{BatyginMorbidelli2013AA}, we showed how resonant dynamics can be represented in an intuitive geometric way, and demonstrated that divergent resonant encounters (where capture necessarily fails) leave the system in a persistent apsidally anti-aligned state. Using the same formalism, \citet{Deck2013} considered the onset of chaotic motion in the unrestricted elliptic three-body problem, and showed that the first order resonance overlap criterion is roughly independent of the planet-planet mass ratio. This is perfectly congruent with our finding that the process of resonant capture and the associated adiabatic threshold only exhibit strong dependence on the ratio of the cumulative secondary mass to the primary mass. 

There is a number of ways in which our theory can be expanded upon. Accordingly, we wish to conclude the paper with a list of possible directions for future development. \\
$\bullet$ To obtain a better relationship between the specified orbital conditions far away from resonance and those at the resonant encounter, it may be fruitful to characterize the pre-encounter secular evolution. This may better inform the true value of the action $\mathcal{J}$ at the time of the appearance of the separatrix. \\
$\bullet$ A more complete version of the model may incorporate the effects of external eccentricity damping or excitation (e.g. \citealt{LeePeale2002,GoldreichSchlichting2014}). Direct modulation of the eccentricities will lead to corresponding changes in the value of the action and may cause $\mathcal{J}$ to shift from a value smaller than $\mathcal{J}_{\rm{C}}$ to a value that exceeds it, or vice-verse. It is possible that the inclusion of this effect will not affect the formulation of capture probabilities because one can envision defining time-dependent canonical coordinates, in which $\mathcal{J}$ is conserved by construction and the evolution rate of $\delta$ is correspondingly accelerated or diminished \citep{Henrard1993}. However, this assertion deserves to be tested explicitly. \\
$\bullet$ In our analysis of probabilistic capture, we have adopted the conventional approach of considering the crossing of a well-defined separatrix, and thus neglected the effects of chaos. In fact, retention of higher order terms in eccentricity and inclination will increase the number of degrees of freedom, rendering the Hamiltonian non-integrable. In a system with multiple degrees of freedom, the vicinity of the separatrix may be engulfed in a chaotic layer, which may alter the outcome of resonant encounters (attempts at characterization of crossing of a stochastic layer have been previously made by \citealt{1993PhyD...68..187H}). \\
$\bullet$ At present, resonant capture theory does not account for external stochastic forces. Such effects have however been shown to compromise the locking and longevity of resonances \citep{Rein2009,Paardekooper2013}. A more complete theoretical framework for this reduction in capture probability may perhaps be constructed with the aid of stochastic calculus (see e.g. \citealt{Adams2008,Batygin2015} for a related discussion). \\
$\bullet$ A somewhat more approximate (but similar in spirit to what has been done here) integrable model for second-order resonances has recently been presented by \citet{Delisle2014}. Accordingly, the existing capture theory for second order resonances in the restricted problem \citep{Henrard1982,BorderiesGoldreich1984} can be adopted for the elliptic problem as well.

Cumulatively, it is clear that the extent of theoretical expansion that can potentially be undertaken is substantial. However, developments such as those proposed above will surely contribute to the construction of a comprehensive model for planetary system formation and evolution, and are thus well-motivated.\\

\textbf{Acknowledgments}. I am thankful to Katherine Deck, Chris Spalding, Peter Goldreich, Greg Laughlin, Heather Knutson, Geoff Blake, Renu Malhotra and Mike Brown for inspirational conversations, as well as to Alessandro Morbidelli, whose careful review of the manuscript and insightful suggestions led to a substantial improvement of the paper.

\end{document}